\def\nlff{{\em nlff}}
\def\omegape{$\omega_{pe}$}
\def\omegae{$\Omega_e$}
\def\ratio{\omegape/\omegae}

\documentclass{aa}  
\usepackage{natbib}
\usepackage{graphicx}
\usepackage{txfonts}
\begin{document}

%
   \title{A new approach to the maser emission in the solar corona}

   \author{S. R\'egnier
          \inst{1}
          }

   \offprints{S. R\'egnier}

   \institute{Department of Mathematics and Information Sciences, 
        Faculty of Engineering and Environment, Northumbria University, 
        NE1 8ST, Newcastle Upon Tyne, UK\\
              \email{stephane.regnier@northumbria.ac.uk}
             }

   \date{Received ; accepted }

 
  \abstract
   {}
   {The electron plasma frequency $\omega_{pe}$ and electron
gyrofrequency $\Omega_e$ are two parameters that allow us to describe
the properties of a plasma and to constrain the physical phenomena at
play, for instance, whether a maser instability develops. 
In this paper, we aim to show that the maser instability can exist in
the solar corona.}
   {We perform an in-depth analysis of the \ratio\ ratio for simple theoretical and complex solar magnetic field
configurations. Using the combination of force-free models for the
magnetic field and hydrostatic models for the plasma  properties, we
determine the ratio of the plasma frequency to the gyrofrequency for
electrons. For the sake of comparison, we compute the ratio for
bipolar magnetic fields containing a twisted flux bundle, and for
four different observed active regions. We also study how \ratio\ is
affected by the potential and non-linear force-free field models.}
   {We demonstrate that the ratio of the plasma frequency to the
gyrofrequency for electrons can be estimated by this novel method
combining magnetic field extrapolation techniques and hydrodynamic
models. Even if statistically not significant, values of \ratio\
$\leq$ 1 are present in all examples, and are located in the low
corona near to photosphere below one pressure scale-height and/or in
the vicinity of twisted flux bundles. The values of \ratio\  are
lower for non-linear force-free fields than potential fields, thus
increasing the possibility of maser instability in the corona.}
   {From this new approach for estimating \ratio, we conclude that
the electron maser instability can exist in the solar corona above
active regions. The importance of the maser instability in coronal
active regions depends on the complexity and topology of the magnetic
field configurations.}


   \maketitle

\section{Introduction}
\label{sec:intro}

Loss-cone driven instabilities play an important role in space
plasmas. The emission of non-thermal radiation is still a puzzling
process despite the accurate observations made in other fields, such
as  planetary magnetospheres \citep{tre06}. This emission is often
explained by the cyclotron maser mechanism for which the $X$-mode and
its harmonics provide the escaping radiation. The growth rate of the
$X$-modes depends on the value of the ratio  of the electron plasma
frequency to the electron gyrofrequency (hereafter $\Xi_e$).
\cite{sha84} studied the development of the $X$-modes and $O$-modes
under the physical conditions of plasmas in the low corona. The study
has been performed for values of $\Xi_e$ less than 2.5. The authors
showed that (i) the electron maser instability dominates the emission
for $\Xi_e \leq 1$ over other loss-cone driven instabilities, and
(ii) the first harmonic of the $X$-mode dominates for $\Xi_e < 0.35$.
In solar physics, the electron maser instability has been studied to
explain the spiky radio burst observed during flares
\citep{hol80,mel82}. Similar studies have revealed the importance of
the $\Xi_e$ thresholds at 0.35 and 1 for the growth of $X$-modes
\citep[e.g.][]{mel84,win85,voc04,tan09,lee13}. The question arising
from these studies is,  {\em Can electron cyclotron maser
emission be a viable mechanism in the solar corona?} To tackle this
question, we  estimate the ratio $\Xi_e$ in the low corona by
assuming that the corona is in equilibrium and thus can be described
by a force-free equilibrium for the magnetic field and a hydrostatic
equilibrium for the plasma parameters.

\begin{figure*}[!ht]
\includegraphics[width=0.498\linewidth, bb = 60 0 504 360]
       {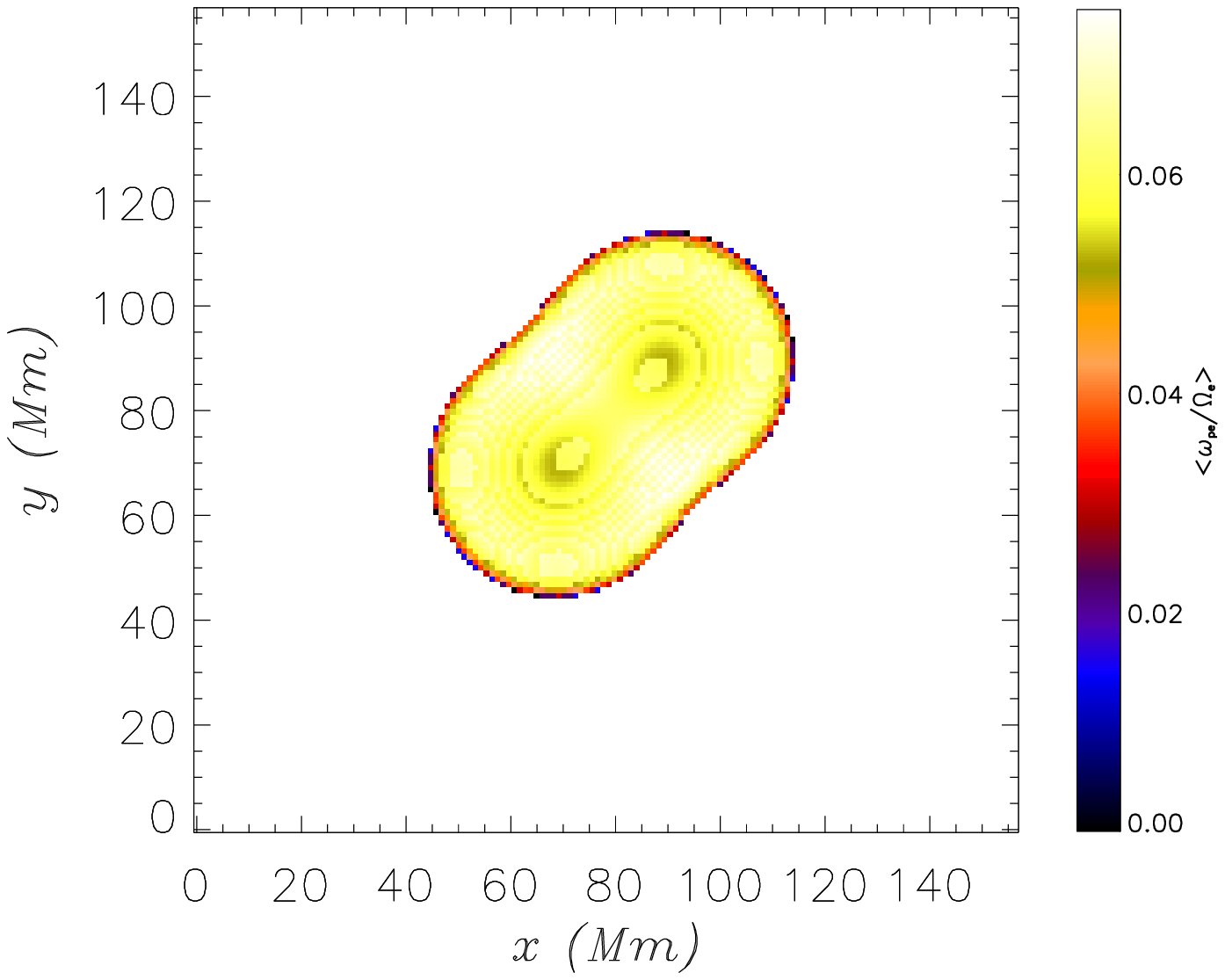}
\includegraphics[width=0.498\linewidth, bb = 60 0 504 360]
       {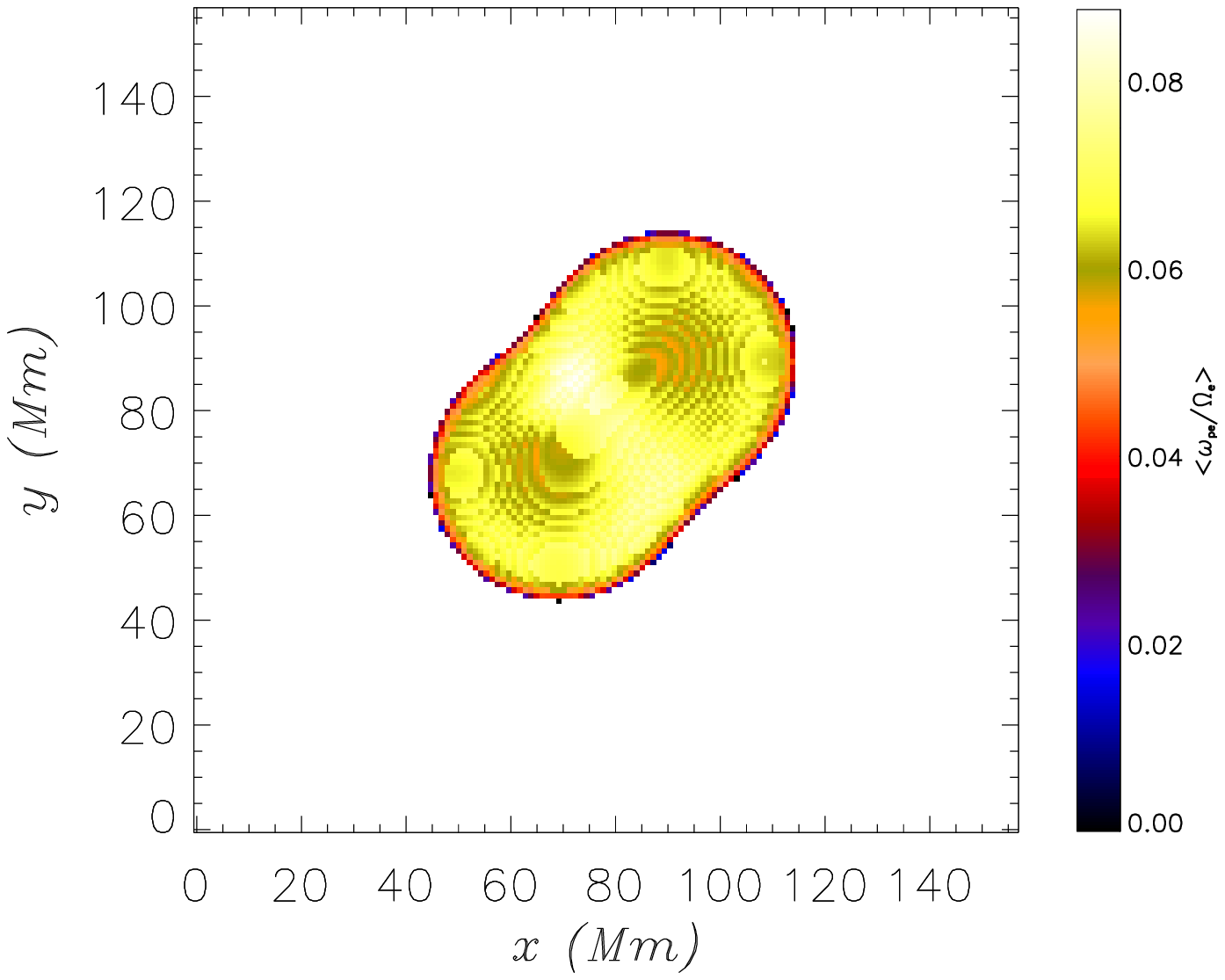}
\hfill
\includegraphics[width=0.498\linewidth, bb = 60 0 504 360]
       {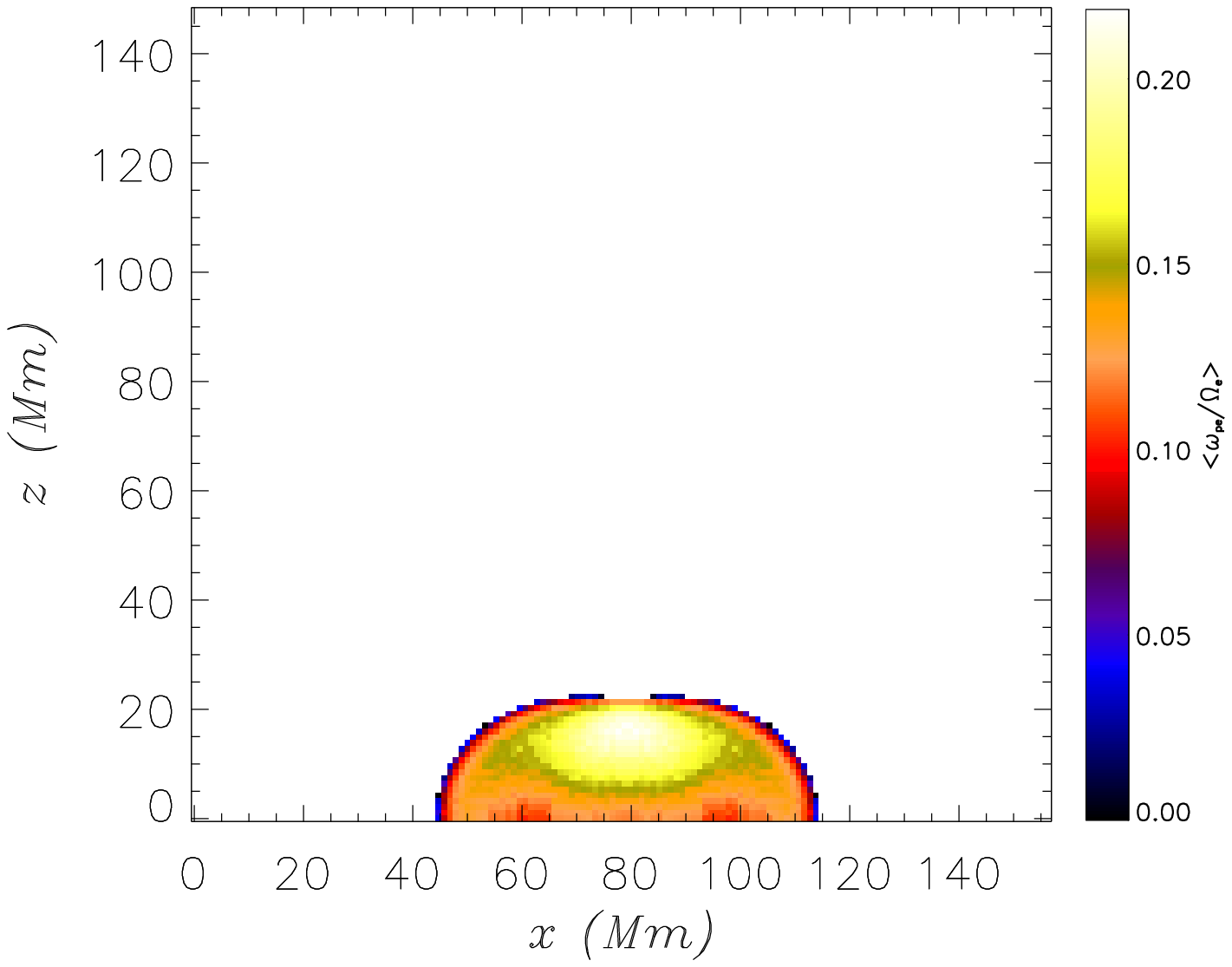}
\includegraphics[width=0.498\linewidth, bb = 60 0 504 360]
       {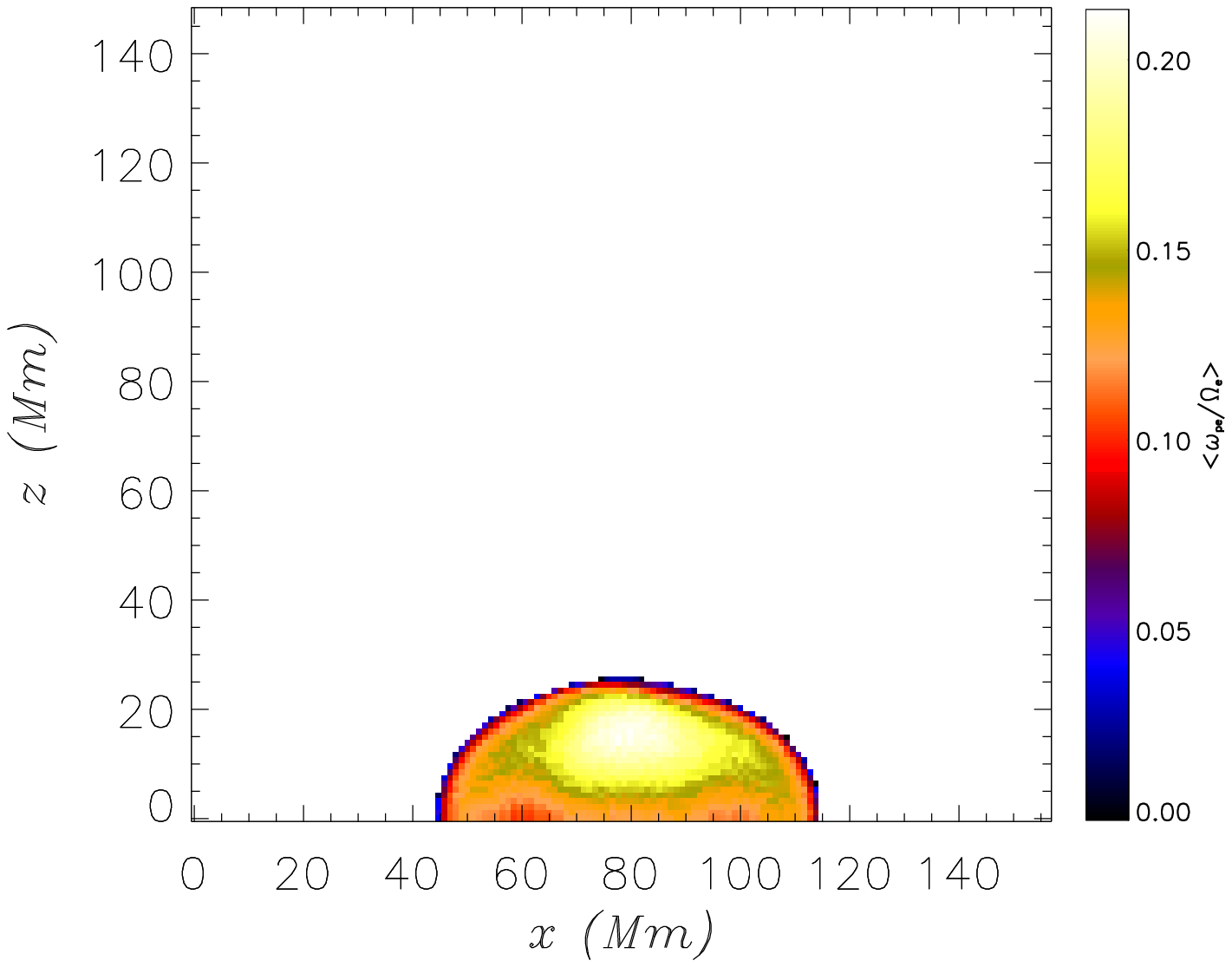}
\caption{Spatial distribution of $\Xi_e \leq 1$ for the bipolar field. (top row) top view; (bottom row) side view for the potential (left) and \nlff\ (right) magnetic fields with varying gravity. The values of $\Xi_e$ are averaged along the 3rd dimension. Only the $\Xi_e \leq 1$ values are colour-coded.}
\label{fig:rng10}
\end{figure*}

The electron population of a plasma can be described by two characteristic frequencies amongst other parameters: the electron plasma frequency, $\omega_{pe}$, and the electron gyrofrequency, $\Omega_e$. The plasma frequency of a cold plasma (neglecting the thermal velocity) is given by
\begin{equation}
\omega_{pe} = \sqrt{\frac{n_e e^2}{m_e \epsilon_0}}
,\end{equation}
where $n_e$ is the electron number density, $e$ is the electric charge
(1.6022$\times$10$^{-19}$ C), $m_e$ is the electron mass (9.1094$\times$10$^{-31}$ kg), and $\epsilon_0$ is the permittivity of free space (8.8542$\times$10$^{-12}$ F m$^{-1}$). The gyrofrequency of electrons is given by
\begin{equation}
\Omega_e = \frac{eB}{m_e}
,\end{equation}
where $B$ is the magnetic field strength. The ratio of the two frequencies,
$\Xi_e = \omega_{pe} / \Omega_e$, can thus be expressed as the ratio of the speed of light to the Alfv\'en speed,
\begin{equation}
\Xi_e = \frac{1}{1.14R} \frac{c}{v_A}
,\end{equation}
where $R$ is the mass ratio between protons and electrons ($R =
1836$) and the coefficient $1.14$ is obtained for a coronal plasma
with a mean atomic weight of $\tilde{\mu} = 0.6$. Thus, knowing the
Alfv\'en speed from the low-order magnetohydrostatic model developed
by \citeauthor{reg08b} (\citeyear{reg08b}), we derive $\Xi_e$ for
different magnetic field configurations. This new method is then used
to determine whether the maser instability is a viable mechanism in
the solar corona.

The paper is organised as follows: after summarising the magnetic field and hydrostatic models in Section \ref{sec:model}, we present the estimate of \ratio\ for a simple bipolar configuration in Section \ref{sec:bip}, and for four different active regions in Section \ref{sec:ar}. We summarise our findings and conclude in Section \ref{sec:concl}.

\section{Magnetic field and density models for the solar corona}
\label{sec:model}

As in \cite{reg08b}, we combine two models to describe the properties
of magnetised plasma in the solar corona: the coronal magnetic field
is assumed to be a force-free field, and the plasma properties are
derived from a stratified atmosphere in hydrostatic equilibrium. In
the following sections, we briefly summarise the different numerical
models.

                \subsection{Potential and non-linear force-free models}

Magnetic field extrapolations are well-known techniques for describing the 3D nature of the coronal magnetic fields  \citep[see review by][]{reg13a}. Under coronal conditions, the magnetic forces dominate the pressure gradient and gravity, and so we regard the corona above active regions as being well described by the force-free approximation \citep[see e.g. recent reviews by][and references therein]{wie12, reg13a}. Throughout this article, the coronal magnetic configurations are computed from the non-linear force-free (\nlff) approximation based on a vector potential Grad-Rubin (\citeyear{gra58}) method by using the
XTRAPOL code \citep{ama97, ama99b}. The {\em nlff} field is governed by the following equations,
\begin{equation}
\vec \nabla \times \vec B = \alpha \vec B,
\end{equation}
\begin{equation}
\vec B \cdot \vec \nabla \alpha = 0,
\label{eq:alphaline}
\end{equation}
\begin{equation}
\vec \nabla \cdot \vec B = 0,
\label{eq:divb}
\end{equation}
where $\vec B$ is the magnetic field vector in the domain $\Omega$
above the photosphere, $\delta \Omega$;  $\alpha$ is a function of
space defined in Cartesian coordinates as the ratio of the vertical
current density, $J_{z}$; and the vertical magnetic field  component,
$B_{z}$:
\begin{equation}
\alpha = \frac{1}{B_z}~\left( \frac{\partial B_y}{\partial x} - \frac{\partial
B_x}{\partial y} \right).
\end{equation} 
From Eq.~(\ref{eq:alphaline}), $\alpha$ is constant along a field line, but varies across field lines. The full description of the Grad-Rubin iterative scheme can be found in \cite{ama97}, and the vector magnetograms used as boundary conditions for the following examples have been detailed, for instance, in \cite{reg08b}. 

For the sake of comparison, we compute both the potential and \nlff\ fields following the technique developed by
\citeauthor{gra58} (\citeyear{gra58}).

                \subsection{Hydrostatic model}
                
In order to define the thermodynamics parameters of the coronal
plasma, we assume that the corona is an isothermal atmosphere
satisfying the hydrostatic equilibrium 
\begin{equation}
-\vec \nabla p + \rho \vec g = \vec 0,
\end{equation}
where $p$ is the plasma pressure, $\rho$ is the density, and $\vec{g}$ is the gravitational force. In agreement with the Harvard-Smithonian model of the solar atmosphere, we consider the hydrostatic equilibrium to be a reasonable assumption above the photosphere at a height ($z_0$) of about 5 Mm. 
As shown in \cite{reg08b}, it is more appropriate to consider the variation with height of
the gravitational field  in order to satisfy the continuity with the properties of the solar wind. Therefore, we will only consider the case of varying gravity satisfying the following equations for the plasma pressure and density,

\begin{equation}
\centering
\left\{ \begin{array}{c l}
p(z) = p_0 \exp{\left(- \frac{R_{\odot}^2}{H~(R_{\odot} + z_0)} \left( \frac{z -
z_0}{R_{\odot} + z} \right) \right)}, & \quad \textrm{and} \\[0.2cm]
\rho(z) = \rho_0 \exp{\left(- \frac{R_{\odot}^2}{H~(R_{\odot} + z_0)} \left(
\frac{z - z_0}{R_{\odot} + z} \right)\right)} & \quad \textrm{for $g = g(z) =
\frac{g_0~R_{\odot}^2}{(R_{\odot} + z)^2}$,}\\
\end{array}
\right.
\end{equation}
where $H = k_BT/(\tilde{\mu}m_pg_0)$ is the pressure scale-height ($k_B = 1.38~10^{-23}$ J~K$^{-1}$, $\tilde{\mu} = 0.6$ for a fully ionised coronal plasma, $m_p = 1.67~10^{-27}$ kg, and $g_0 = g(R_{\odot}) = 274$
m~s$^{-2}$), and $p_0$ and $\rho_0$ are characteristic values of the plasma pressure and density at $z_0$. The pressure scale-height and the density $\rho_0$ are the two free parameters of the model. Typical values are $H = 50$ Mm giving a coronal temperature of 1 MK and the number density correpsonding to $\rho_0$ is $n_0 = 10^{9}$ cm$^{-3}$. 

For a gravitational field varying with height, the Alfv\'en speed is thus given by
\begin{equation}
v_A(x, y, z) = \frac{B(x, y, z)}{\sqrt{\mu_0 \rho_0}} \exp{\left(
\frac{R_{\odot}^2}{2H~(R_{\odot} + z_0)} \left( \frac{z - z_0}{R_{\odot} + z}
\right)\right)},
\label{eq:vagz}
\end{equation}
where $B$ is the magnetic field strength.

\begin{figure}[!h]
\includegraphics[width=1.\linewidth]
       {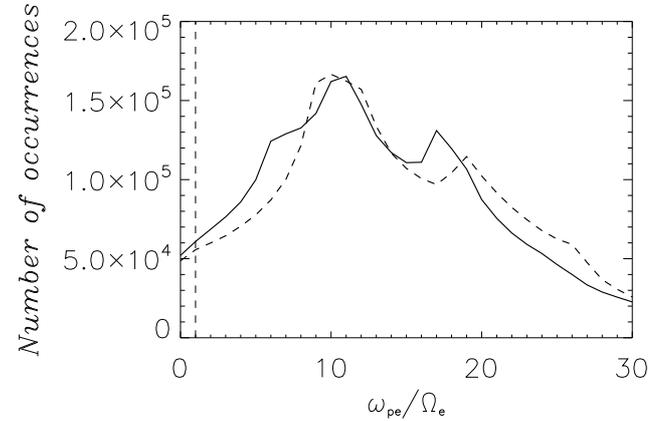}
\caption{Distribution of $\Xi_e$ for the bipolar magnetic field within the entire computational volume (cutoff at $\Xi_e = 30$) for the potential (dashed line) and \nlff\ (solid line) magnetic field models. The vertical dashed black line is the $\Xi_e = 1$ threshold.}
\label{fig:rng10_dist}
\end{figure}

In summary, for all cases (bipolar field, and active regions), we use two different models (potential field and \nlff\ field both with a varying gravity) to analyse the variation of the \ratio\ ratio.
                
\section{Bipolar fields}
\label{sec:bip}

\begin{figure*}[!ht]
\includegraphics[width=0.498\linewidth, bb = 40 0 504 360]
       {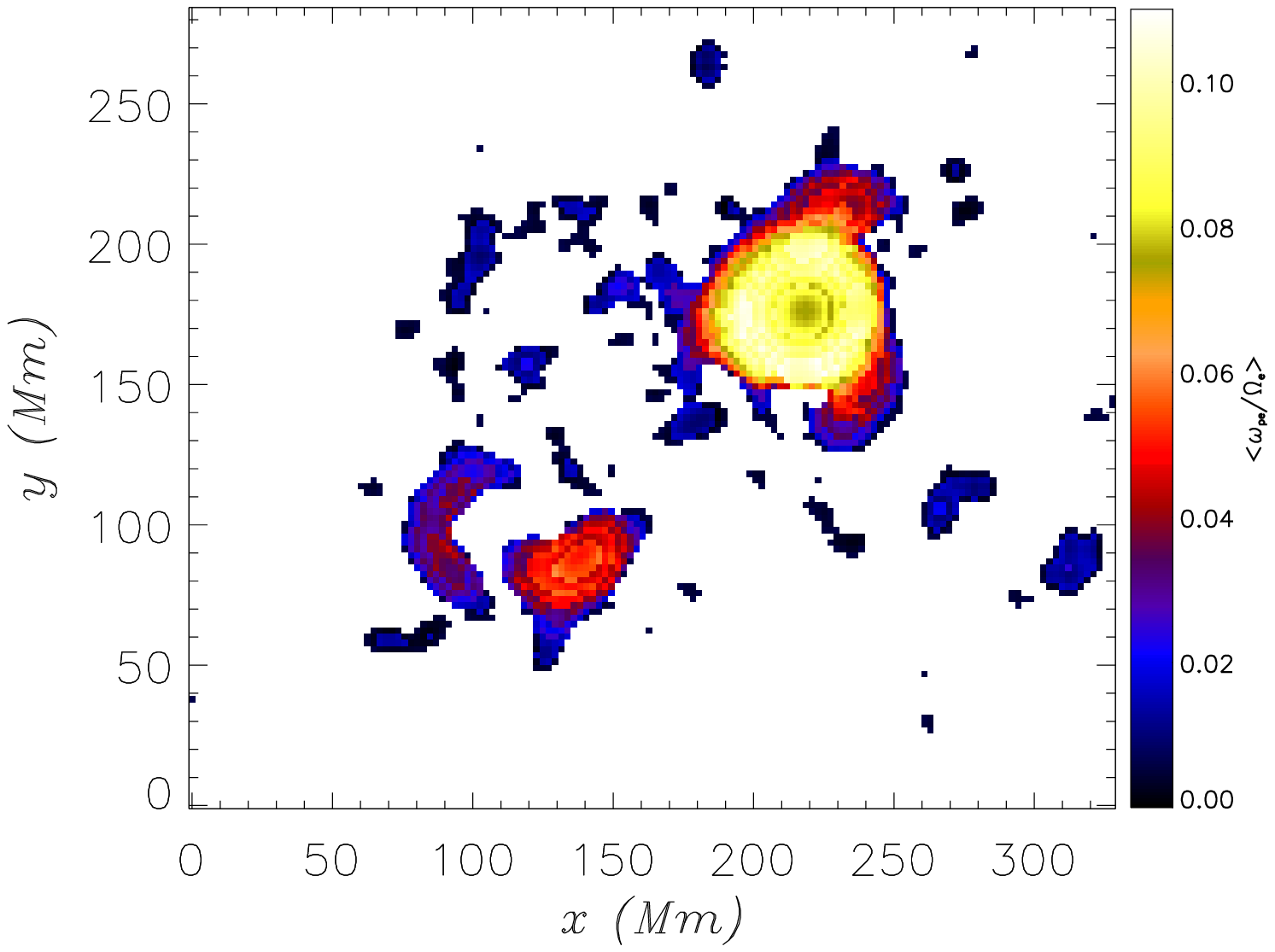}
\includegraphics[width=0.498\linewidth, bb = 40 0 504 360]
       {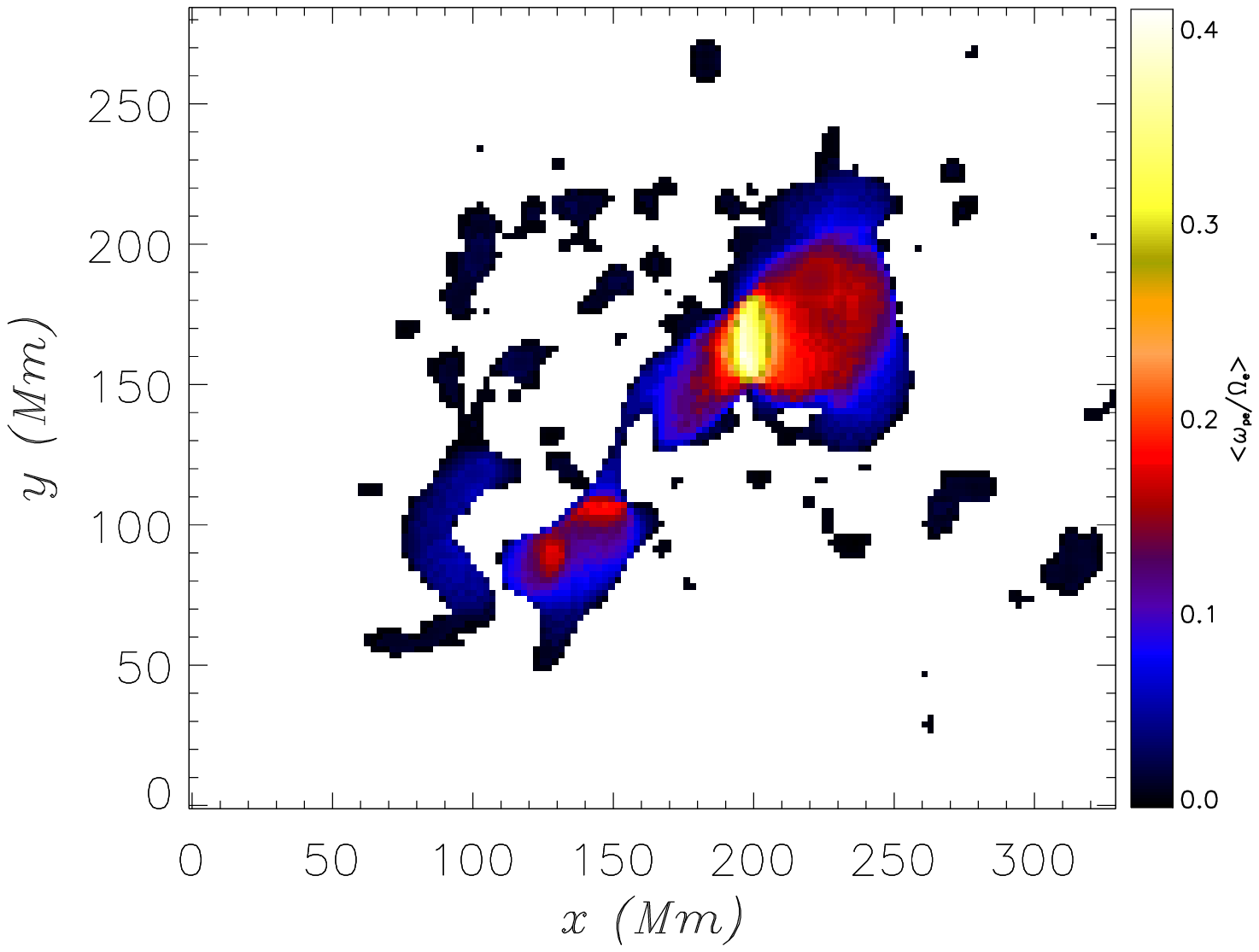}
\includegraphics[width=0.498\linewidth, bb= 10 0 480 270] 
      {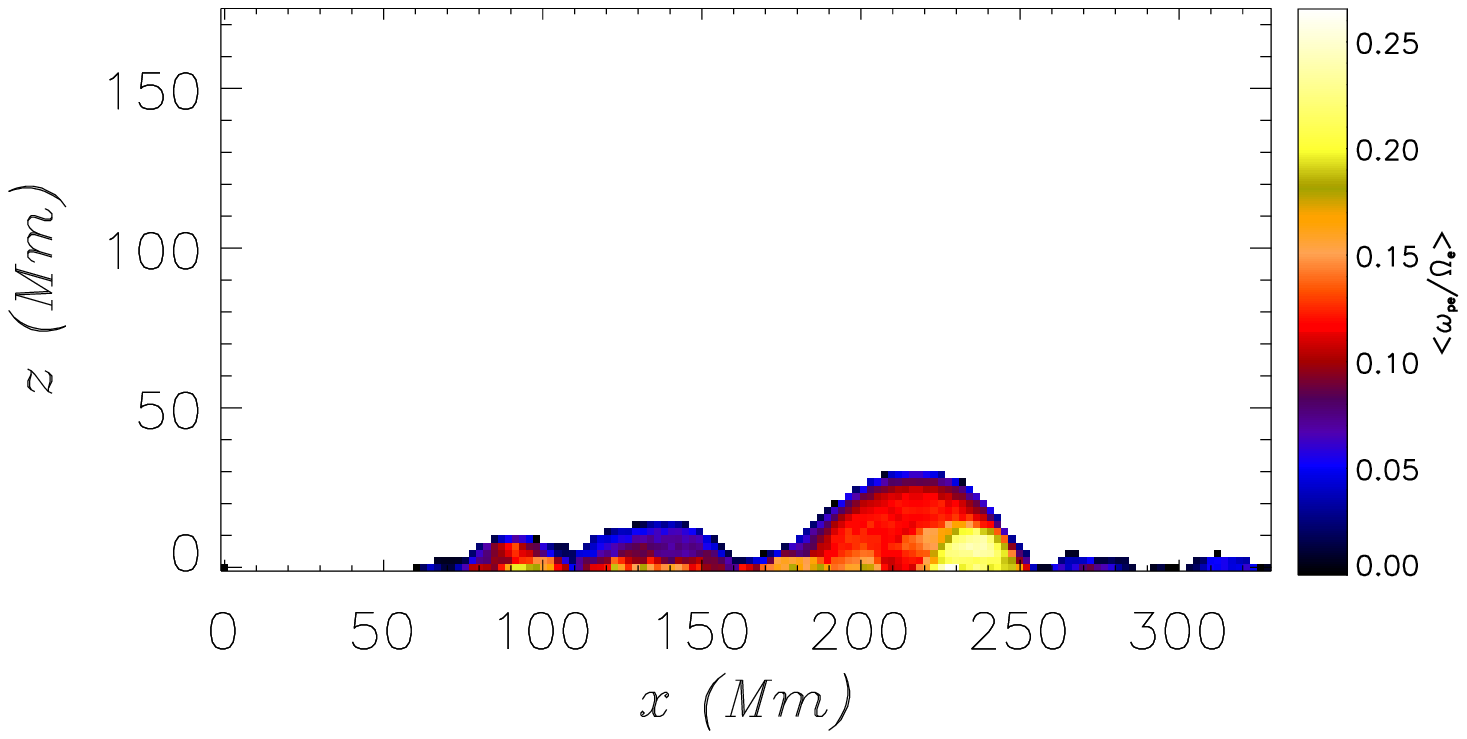}
\includegraphics[width=0.498\linewidth, bb= 10 0 480 270]
       {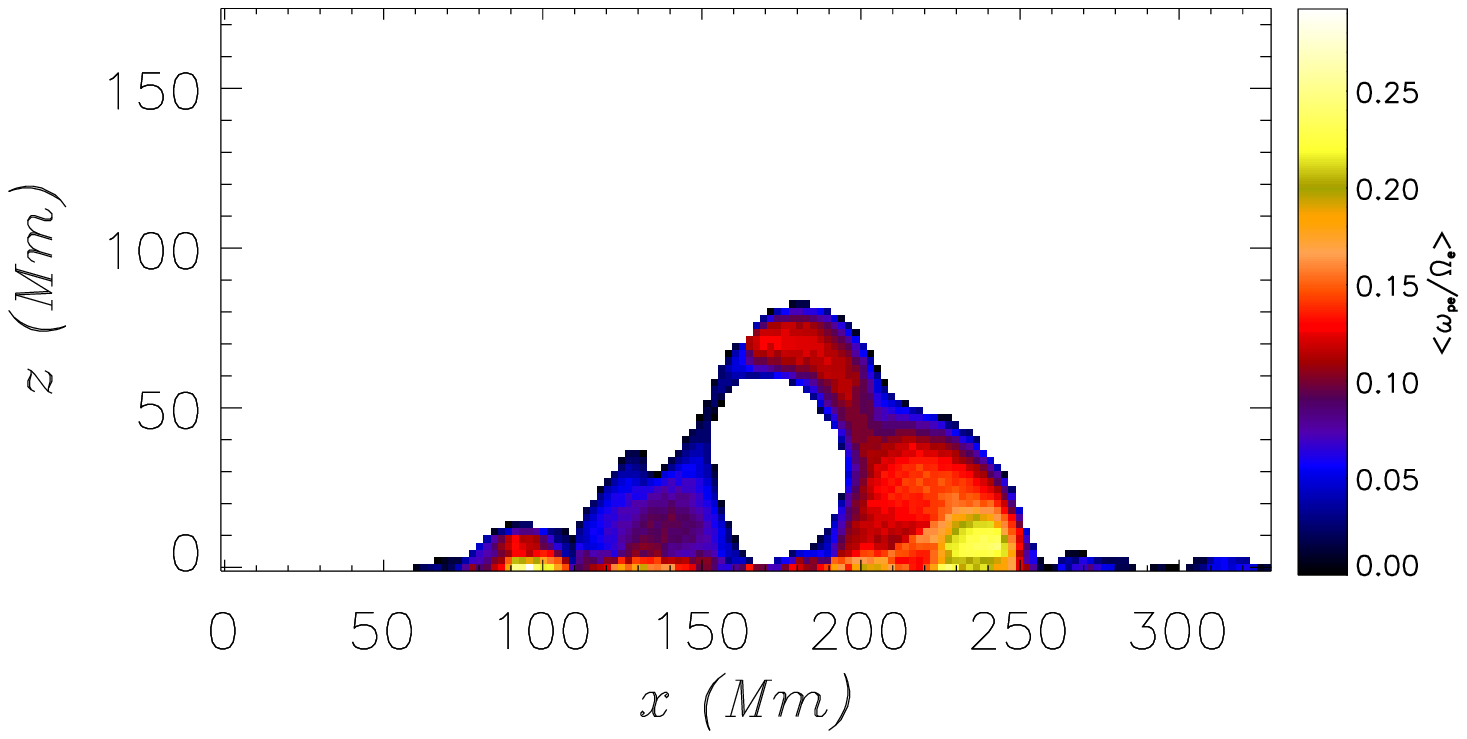}
\caption{Same as Fig.~\ref{fig:rng10} for AR8151}
\label{fig:ar8151}
\end{figure*}

As a test case for the above method, we use a simple bipolar magnetic field distribution on the photosphere. For all models, the vertical component of the magnetic field, $B_z$, is given by a Gaussian distribution. To compute the \nlff\ field, the vertical electric
current density is injected in the magnetic configuration following a
photospheric distribution of the form
\begin{equation} J_z(r) = 2~J_{z0}~[r^2 -
C_0]~\exp{\left(-\frac{r^2}{\sigma^2}\right)}, \label{eq:rng} \end{equation}
where $r$ is measured from the centre of the magnetic polarity and $C_0$ is a constant that ensures a zero net current. This $J_z$ distribution is typically a second-order Hermite polynomial function that allows for return current on the surface of the flux tube (see \citeauthor{reg12} \citeyear{reg12}). The vertical magnetic field strength $B_z$ is 2000 G at the centre of the polarities and $J_z$ has a maximum strength of 10 mA~m$^{-2}$. Both the potential and \nlff\ models are computed on a 148$\times$148$\times$148 computational box with a pixel size of 1 Mm. 

In Fig.~\ref{fig:rng10}, we plot the spatial distribution of $\Xi_e \leq 1$ for the potential and \nlff\ fields with varying gravity (from left to right). The first row shows a top view ($xy$-plane), whilst the second row shows a side view ($xz$-plane). The values of $\Xi_e$ are averaged along the third dimension. The values of $\Xi_e \leq 1$ are localised where the magnetic field strength is strong, below a height of 25 Mm. For this bipolar magnetic field configuration, the values of $\Xi_e$ are similar across both models (see also Table~\ref{tab:xi}). 

The distributions of $\Xi_e$ within the volume are shown for both models in Fig.~\ref{fig:rng10_dist} (values of $\Xi_e$ between 0 and 30). The two distributions have similar shapes with two main peaks around 10 and 15, and a tail for large values of $\Xi_e$ which tends to zero. For $\Xi_e \leq 1$, the distributions have a number of occurrences which is not significant whatever the magnetic field model. We confirm this in  Table~\ref{tab:xi}  by listing the percentage of $\Xi_e$ in given intervals. We use the same intervals as in \cite{sha84}. The percentage of $\Xi_e \leq 1$ is about 1.5\% for all models, whilst $\sim$95\% of $\Xi_e$ values are above 2.5. In this example, the models with varying gravity give similar results whether or not the magnetic configuration contains electric currents. Thus, the maser emission is weakly influenced by the electric currents or the twist and shear in a simple magnetic configuration (without a topology).  

The conclusions from this test case are (i) values of $\Xi_e \leq 1$ exist in coronal-like magnetic field configurations, and (ii) these values are located above large photospheric magnetic field strengths.

\section{Active regions}
\label{sec:ar}

We now analyse the values of \ratio\ for four active regions at different stages of their evolution. 

\paragraph{AR8151 \\}

\begin{figure*}[!ht]
\includegraphics[width=.498\linewidth]{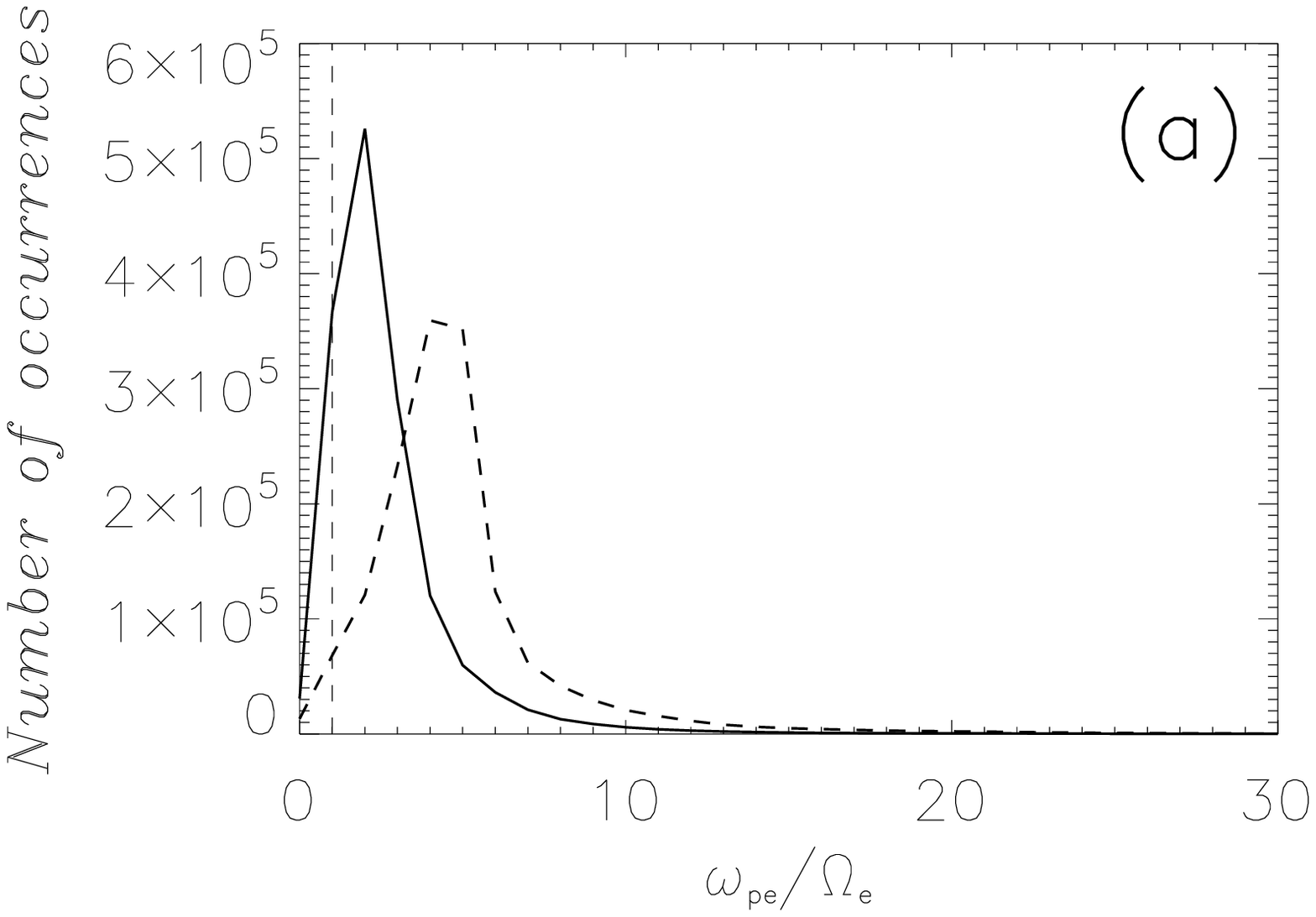}
\includegraphics[width=.498\linewidth]{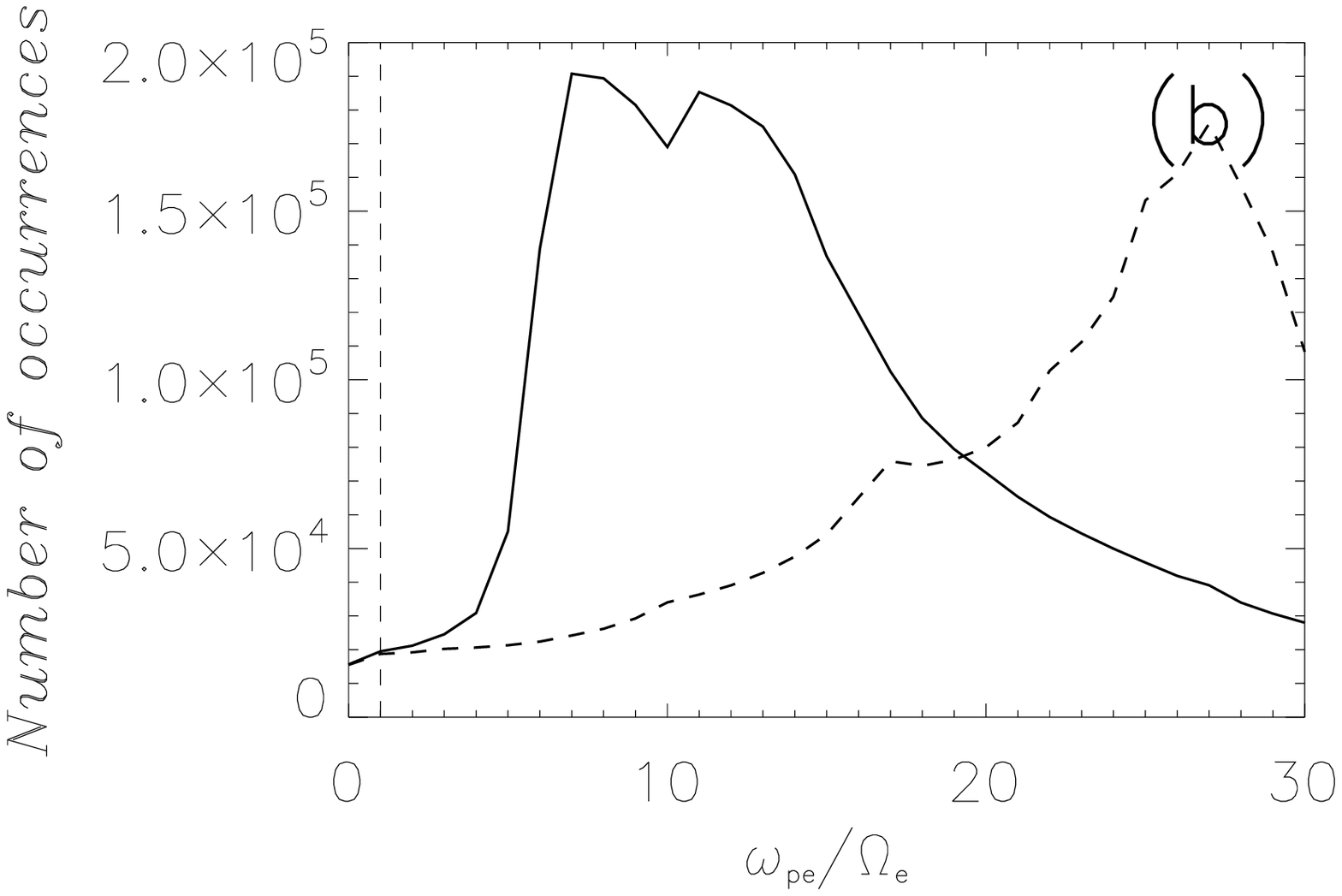}
\includegraphics[width=.498\linewidth]{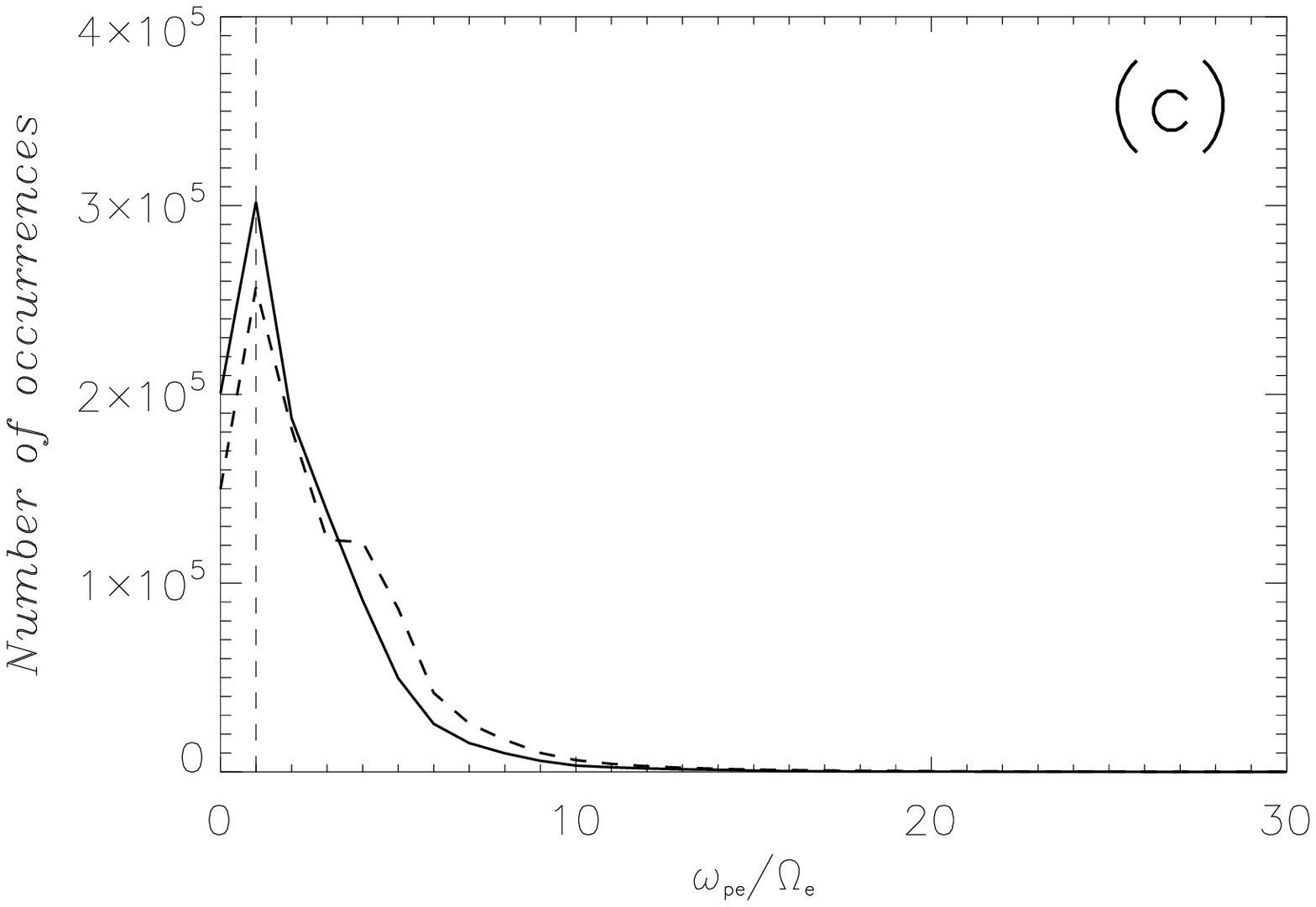}
\includegraphics[width=.498\linewidth]{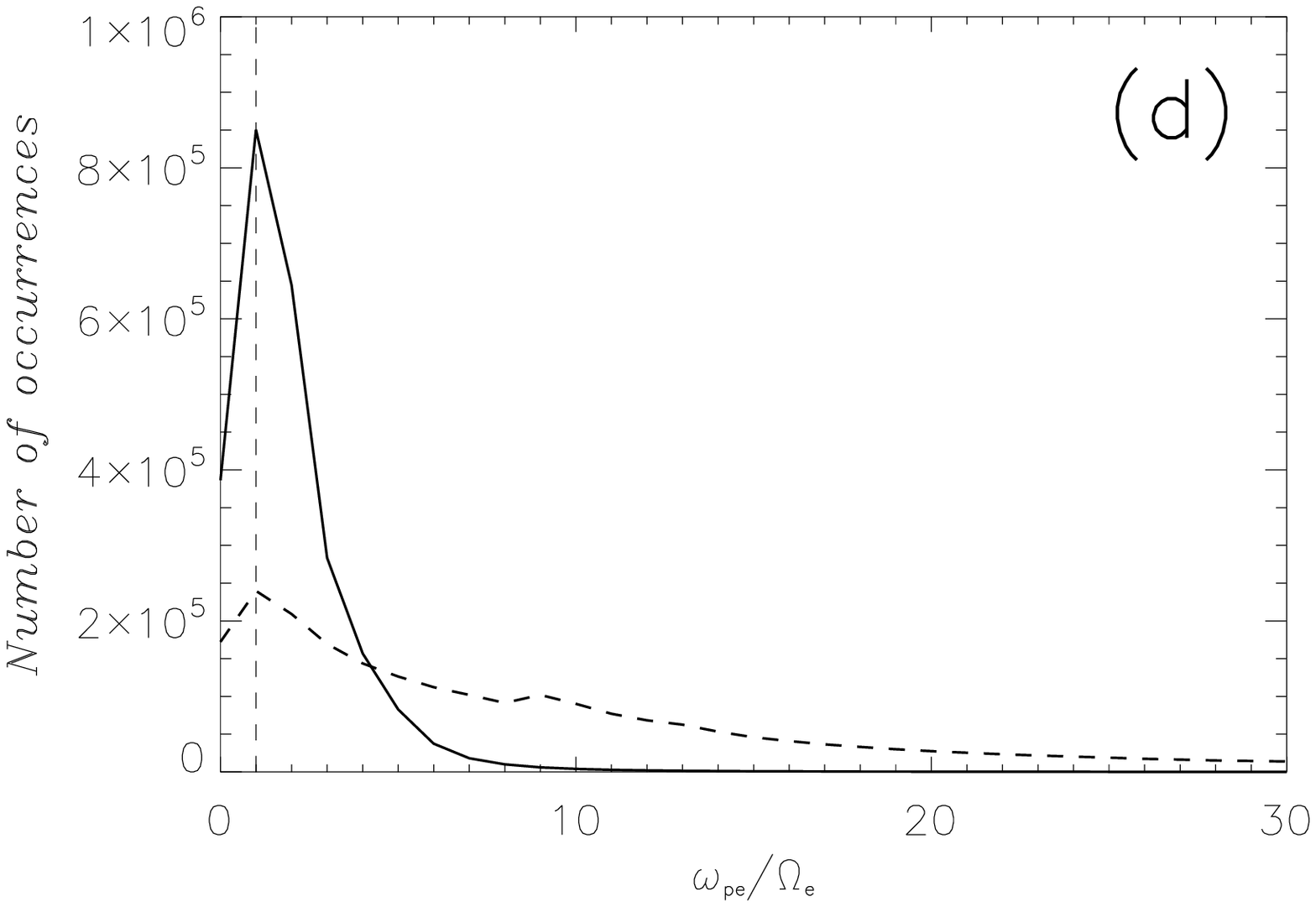}
\caption{Same as Fig.~\ref{fig:rng10_dist} for (a) AR8151, (b) AR8210, (c) AR9077, and (d) AR10486}
\label{fig:ar_dist}
\end{figure*}

The active region AR8151 is an old decaying active region with moderate and highly twisted flux bundles observed as a filament and an X-ray sigmoid, respectively \citep{reg02,reg04}. In Fig.~\ref{fig:ar8151}, the $\Xi_e \leq 1$ values are located where the magnetic field strength is large, i.e. above sunspots and plages. For the potential model, the values of $\Xi_e \leq 1$ are concentrated below 30 Mm, i.e. within one pressure scale-height above the surface. For the \nlff\ model, the spatial distribution shows significant values of $\Xi_e$ above 50 Mm; these values characterise the influence of electric currents present in the magnetic configuration and are associated with highly twisted flux bundles and sheared arcades. The twisted flux bundles existing in this active region generate a local increase in the magnetic field strength, i.e. in the local Alfv\'en speed (see Eq.~\ref{eq:vagz}). As can be seen in Fig.~\ref{fig:ar_dist}a, the distribution of $\Xi_e$ peaks at 2 for the \nlff\ model and at 5 for the potential model. Both distributions are narrower (with a typical width of about 2) than in the magnetic bipole example presented in Section~\ref{sec:bip}. Unlike the bipolar field, there is a clear difference between the potential and \nlff\ models; the $\Xi_e$ values are smaller for the \nlff\ magnetic field. Both distributions  have a tail that becomes statistically insignificant for values of \ratio\ above 10. These results lead us to consider that, for this active region, the \nlff\ model has a different behaviour than the potential field model owing to the presence of twisted flux bundles and the complexity of the magnetic field; a proxy of the complexity of the coronal magnetic field is given by the complexity of the photospheric magnetic field. In terms of statistics, Table~\ref{tab:xi} shows that values of $\Xi_e \leq 1$ represent $\sim$0.8\% for the potential model and $\sim$1.4\% for the \nlff\ model. The main difference is in the percentage of $\Xi_e$ between 1 and 2.5, which reaches 40\% for the \nlff\ field, whilst it is just around 8\% for the potential model; as noticed previously, this is due to the strong electric currents present in the active region, which imply sheared and twisted magnetic field lines leading to the conclusion that more realistic magnetic fields like the \nlff\ model are more favourable to maser  emission.

\paragraph{AR8210 \\}

\begin{figure*}[!ht]
\includegraphics[width=0.498\linewidth, bb= 20 0 480 320]
       {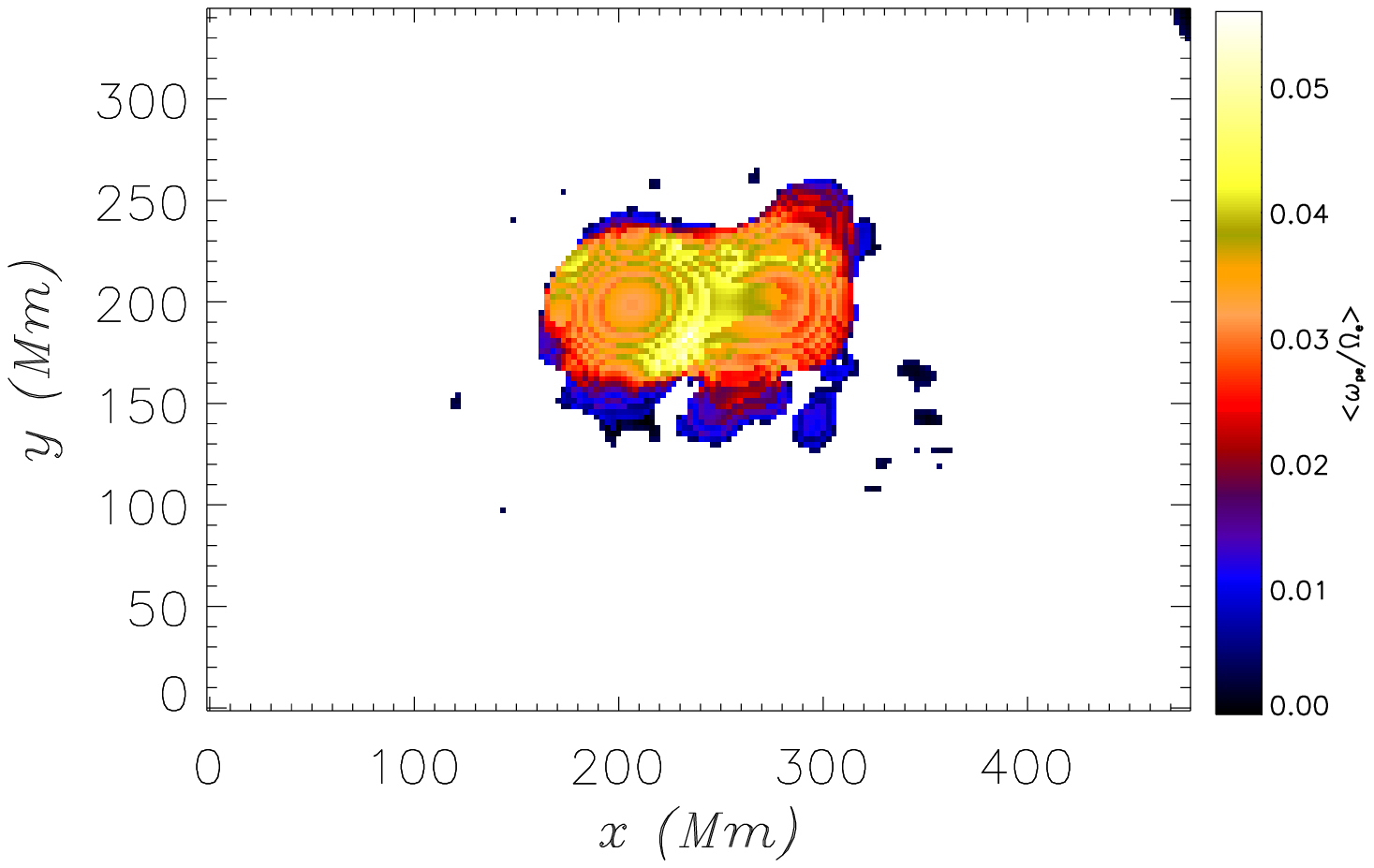}
\includegraphics[width=0.498\linewidth, bb= 20 0 480 320]
       {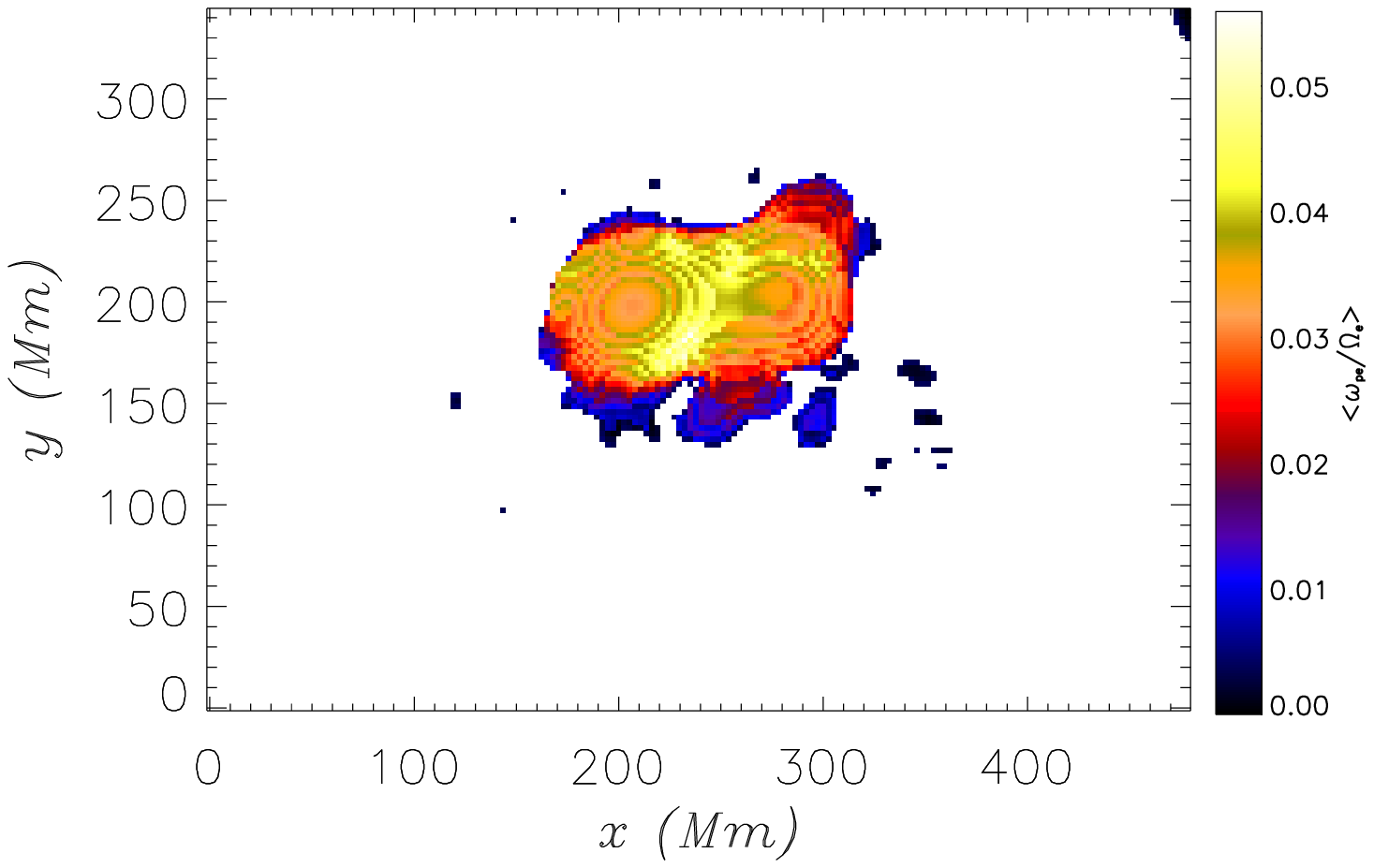}
\includegraphics[width=0.498\linewidth, bb= 20 0 480 330]
       {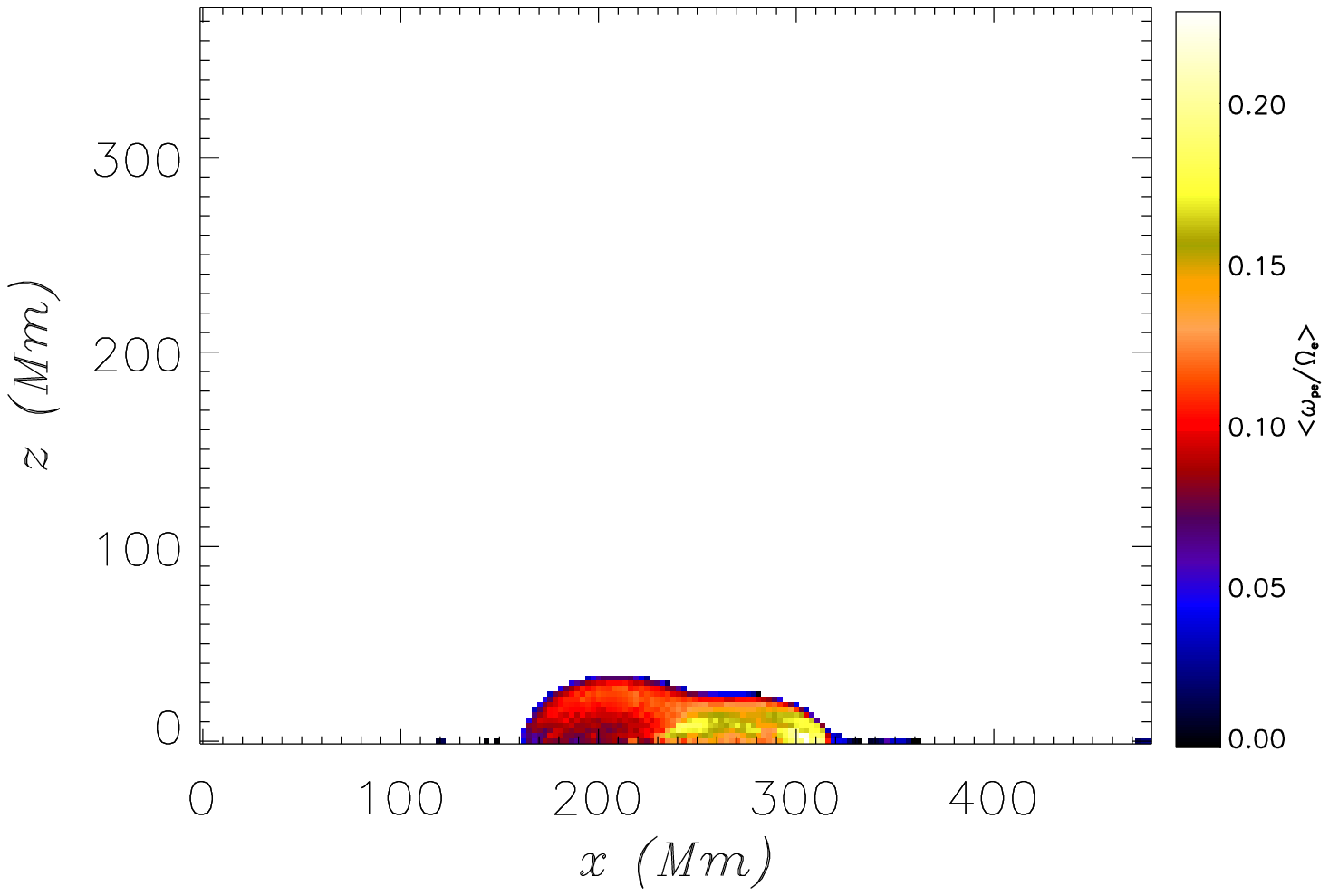}
\includegraphics[width=0.498\linewidth, bb= 20 0 480 330]
       {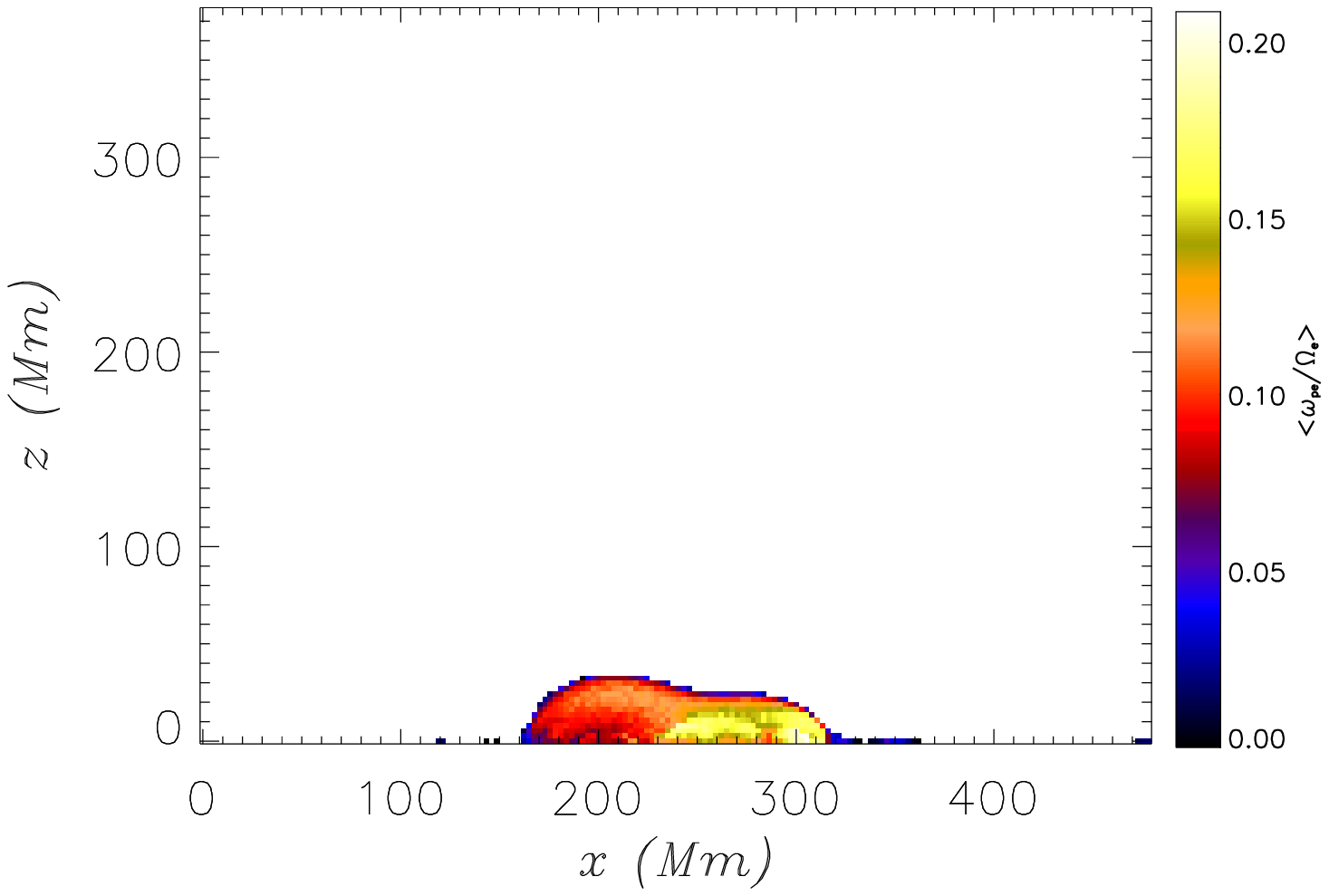}
\caption{Same as Fig.~\ref{fig:rng10} for AR8210}
\label{fig:ar8210}
\end{figure*}

AR8210 is a relatively new active region which has produced numerous flares.
The main features in AR8210 are a clockwise-rotating sunspot surrounded by diffuse
polarities of opposite sign, and a small emerging parasitic polarity interacting
with the pre-existing magnetic topology of the field \citep{reg06}. A time series of \nlff\ magnetic field
extrapolations has shown that the 3D magnetic configuration is close to
potential (no twisted bundles, and small noticeable sheared arcades) and does not contain a large amount of free magnetic energy (estimated to $\Delta E_m = 2.6~10^{31}$ erg), i.e. not sufficient to produce large flares. 

In Fig.~\ref{fig:ar8210}, we plot the spatial distribution of $\Xi_e$ within the whole
computational volume. As for the bipolar field described in
Section~\ref{sec:bip}, the values of $\Xi_e \leq 1$ are located just above the
strong magnetic field regions and do not extend above 30 Mm. In Fig.~\ref{fig:ar_dist}b, the distributions of $\Xi_e$ indicate
that the significant values of the ratio are above 3. As noted for AR8151, the \nlff\ model decreases significantly the values of $\Xi_e$ compared to
the potential field model. The $\Xi_e$ distribution peaks between 7 and 13 for the \nlff\ model, whilst it peaks at 27 for the potential model. In terms of statistics (see
Table~\ref{tab:xi}), the percentage of $\Xi_e \leq 1$ is less than 0.5\% for both
models, and the percentage of $\Xi_e \geq 2.5$ is about 98\%. These results are consistent with the fact that no twisted flux bundles or highly sheared arcades have been found in the 3D extrapolated magnetic field \citep{reg06}.

\paragraph{AR9077 \\}

\begin{figure*}[!ht]
\includegraphics[width=0.498\linewidth, bb= 50 0 500 360]
       {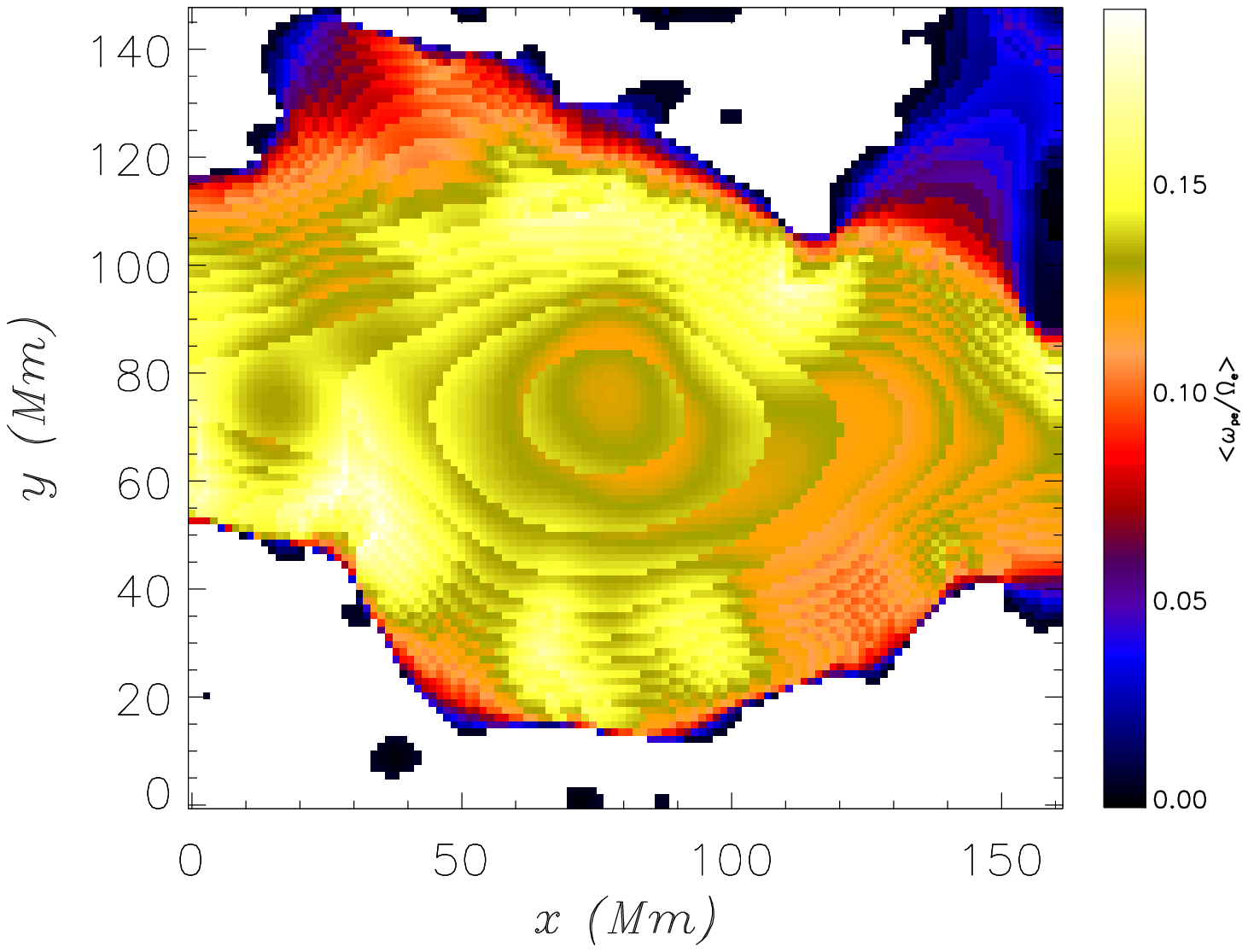}
\includegraphics[width=0.498\linewidth, bb= 50 0 500 360]
       {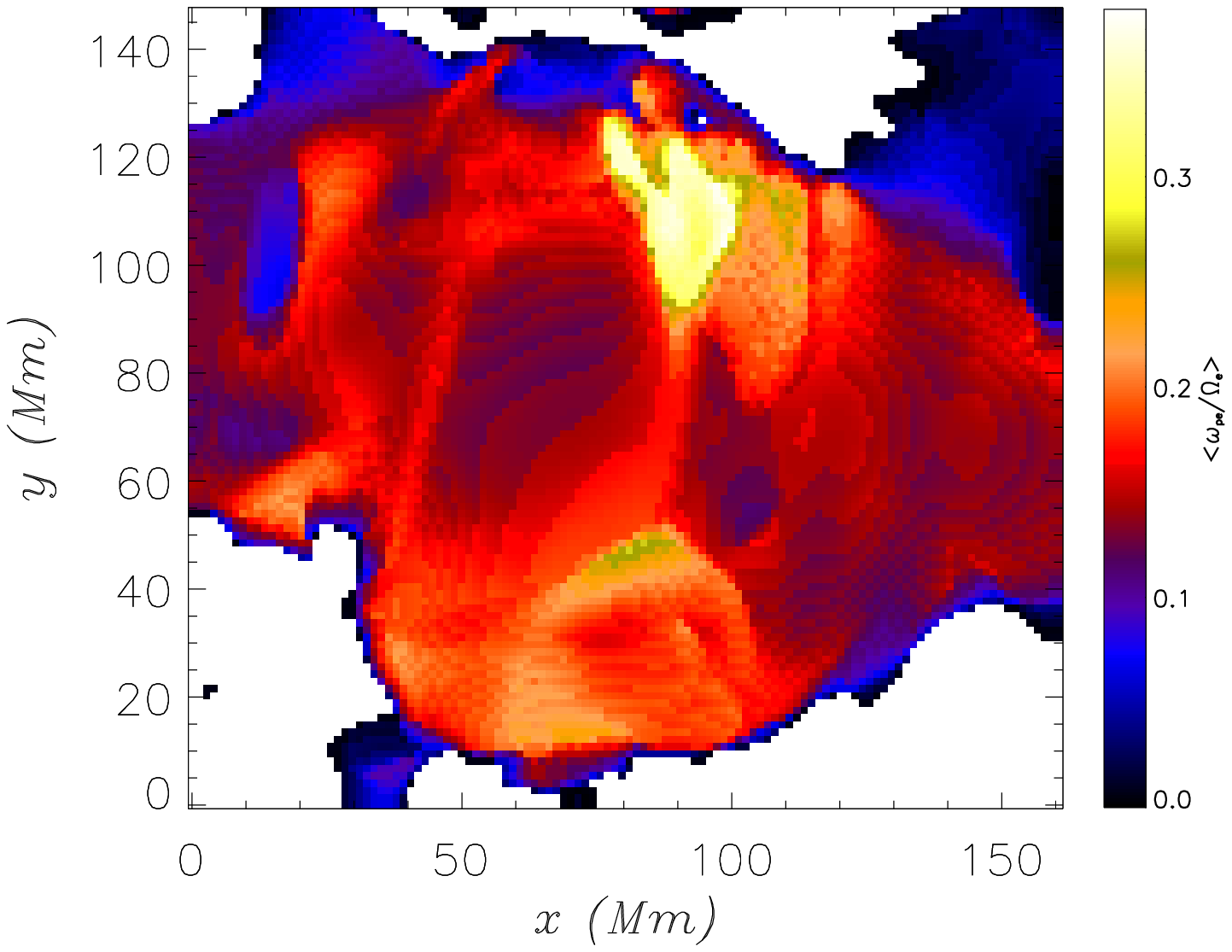}
\includegraphics[width=0.498\linewidth, bb= 10 10 470 300]
       {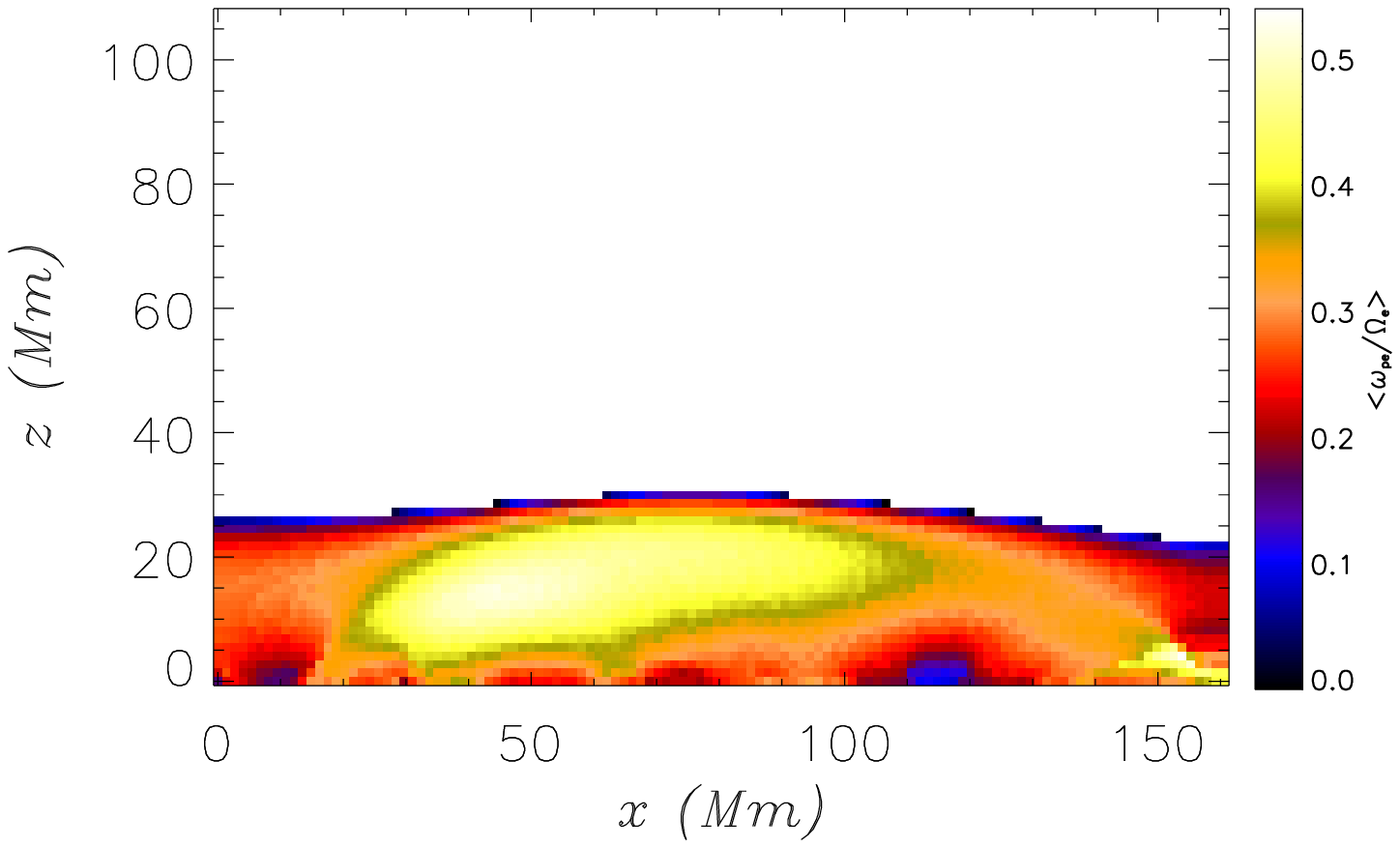}
\includegraphics[width=0.498\linewidth, bb= 10 10 470 300]
       {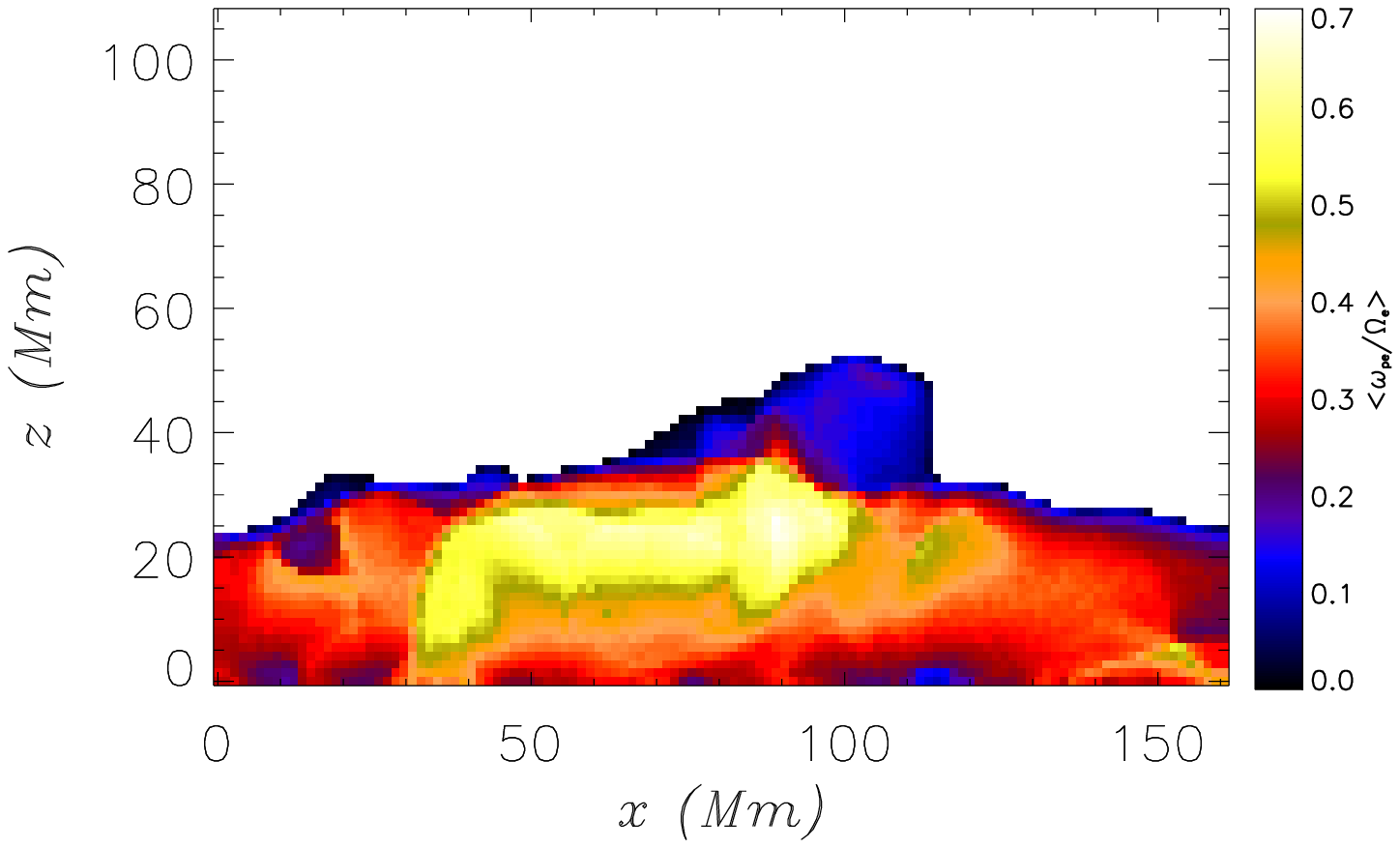}
\caption{Same as Fig.~\ref{fig:rng10} for AR9077}
\label{fig:ar9077}
\end{figure*}

AR9077  produced a X5.7 flare on 2000 July 14 known as the Bastille
Day flare \citep[e.g.][]{kos01,wan04}. The magnetic field
extrapolation was performed after the occurrence of the X-class flare
and thus describes the relaxation phase associated with post-flare
loops. It has been shown that the magnetic field configuration is
quite close to a potential field with a small amount of shear and
twist. These magnetic field properties are reflected in the spatial
distribution of $\Xi_e$ (Fig.~\ref{fig:ar9077}): the $\Xi_e \leq 1$
values are concentrated above regions of large magnetic field
strengths as was seen for the previous examples. Both potential and
\nlff\ fields have similar spatial distributions below 30 Mm.
Futhermore, the \nlff\ field exhibits an excess of low $\Xi_e$ values
above 30 Mm indicating the presence of twisted/sheared magnetic field
lines increasing locally the magnetic field strength. Compared to
AR8151 (see Fig.~\ref{fig:ar8151}), the volume containing the $\Xi_e
\leq 1$ values above 30 Mm is small, and does indicate a small amount
of twist or shear in the magnetic configuration. In
Fig.~\ref{fig:ar_dist}c, the distributions for both models peak at
about 1 with a percentage of values with $\Xi_e \leq 1$ between 14\%
and 19\% (see Table~\ref{tab:xi}). It reinforces the fact that the
magnetic configuration of AR9077 at this stage of its evolution is
close to potential, although the number of locations where the maser
instability could occur is large (about 20\%), filling  the volume
below 30 Mm almost entirely.   

\paragraph{AR10486 \\}

\begin{figure*}[!ht]
\includegraphics[width=0.498\linewidth, bb= 60 0 500 360]
       {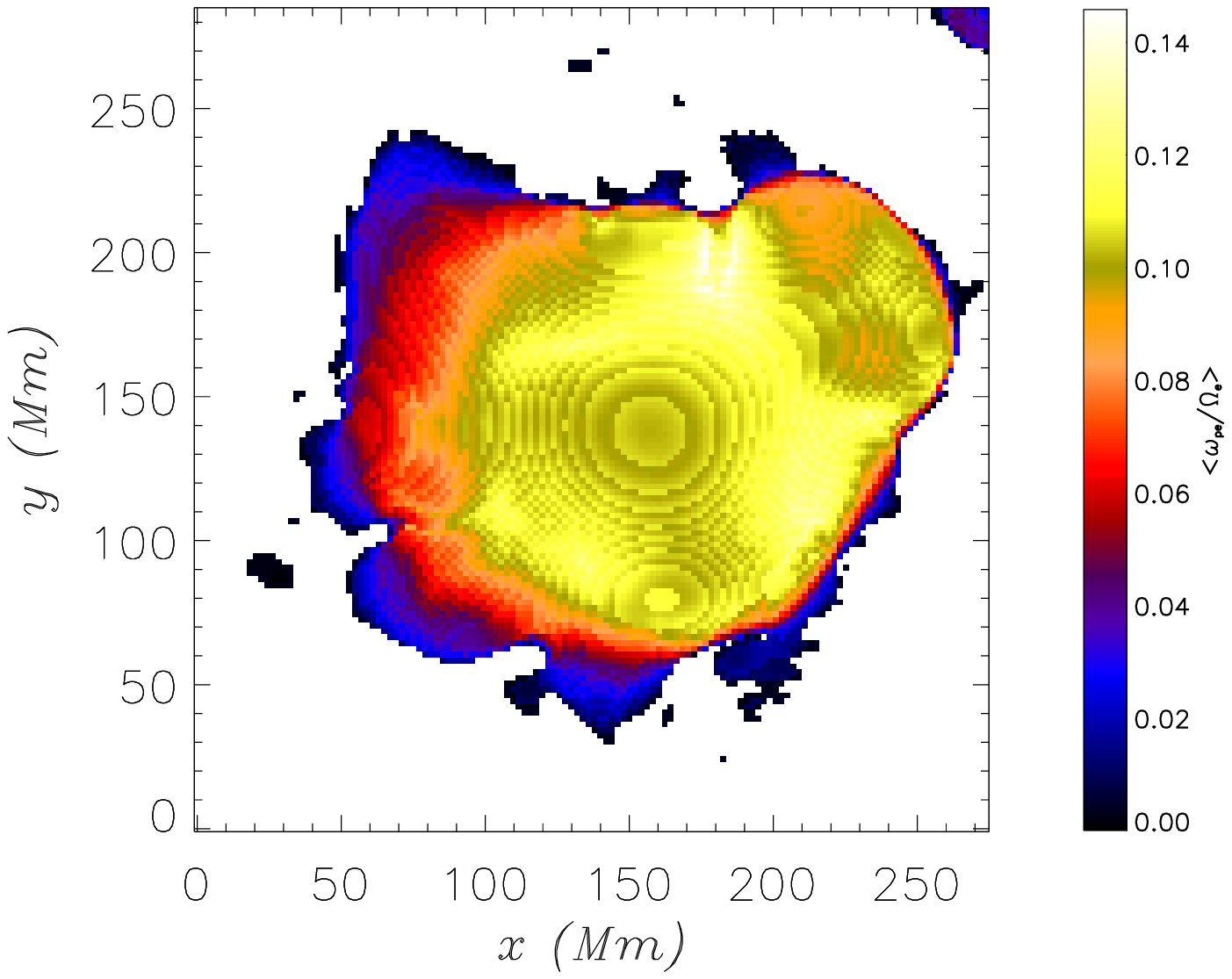}
\includegraphics[width=0.498\linewidth, bb= 60 0 500 360]
       {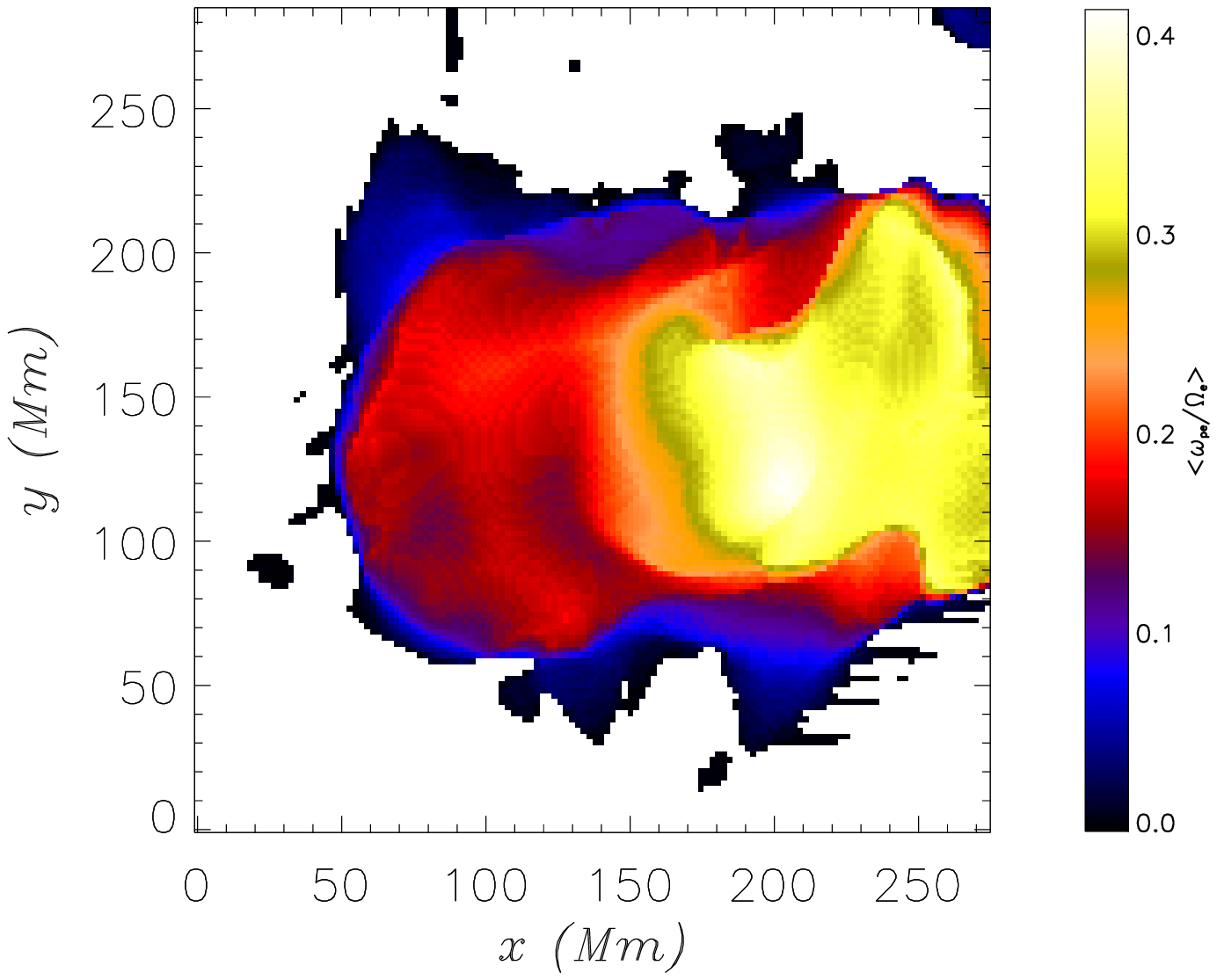}
\includegraphics[width=0.498\linewidth, bb= 0 10 460 360]
       {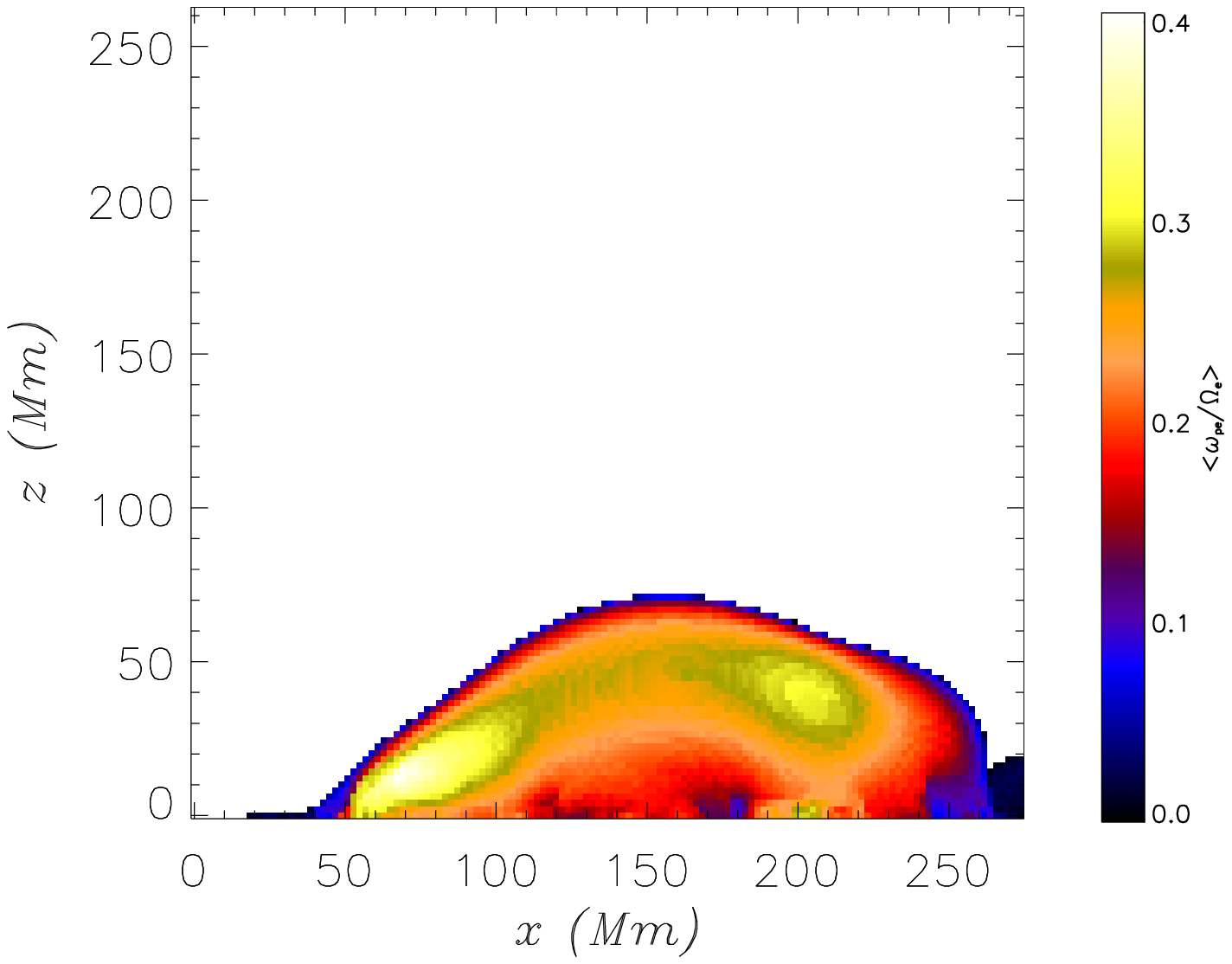}
\includegraphics[width=0.498\linewidth, bb= 0 10 460 360]
       {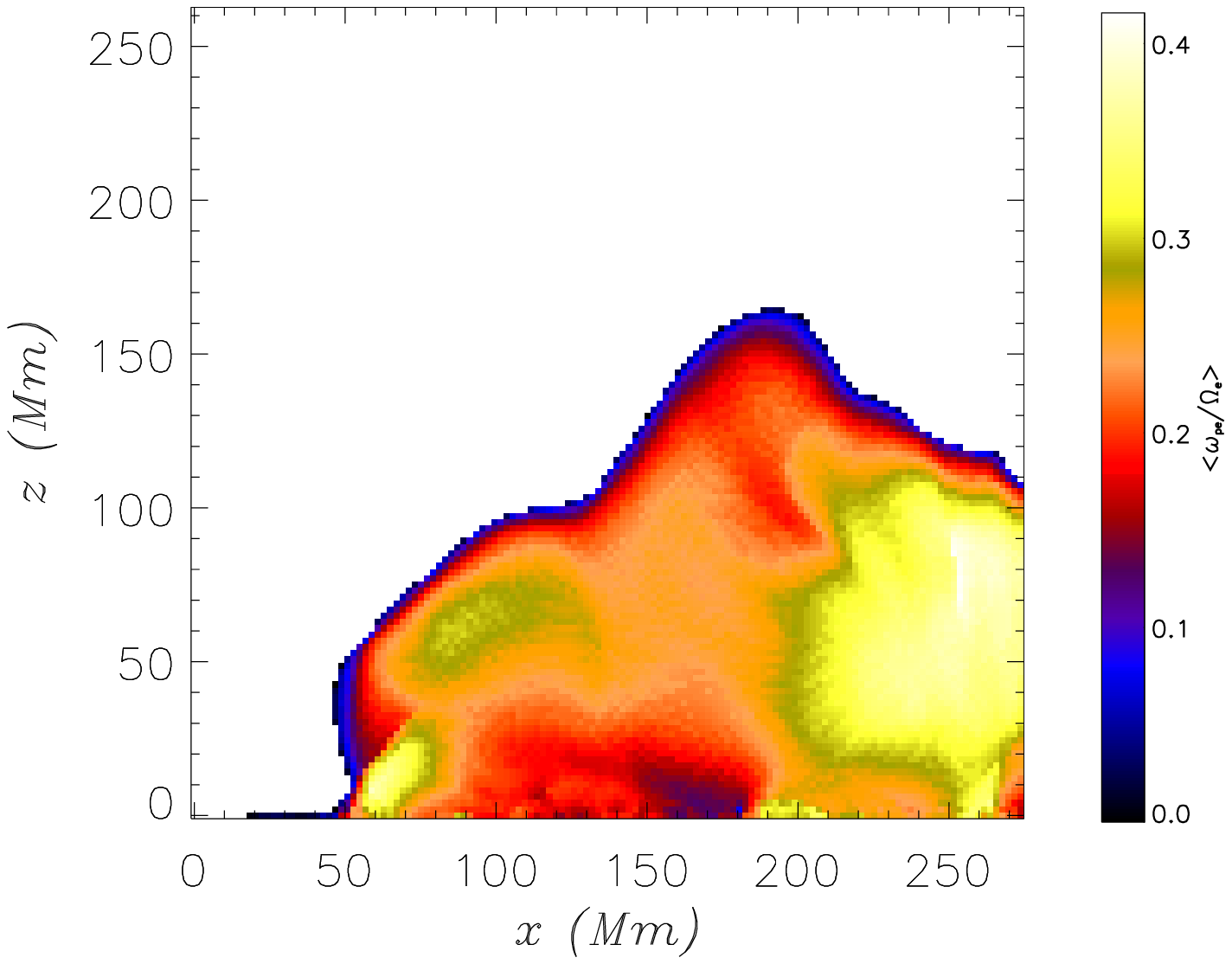}
\caption{Same as Fig.~\ref{fig:rng10} for AR10486}
\label{fig:ar0486}
\end{figure*}

The active region AR10486 is the source of the series of strong
flares in October-November 2003 known as the Halloween events,
including the strongest flare recorded for the corresponding solar
cycle estimated to be a X28 flare on November 4. The magnetic field
configuration presented here is obtained before the occurrence of the
X17 flare located near the disk centre on October 28. For this flare,
the energetics and magnetic topology have been studied in detail
\citep[see e.g.][]{reg05c,man06,zuc09}; the magnetic energy contained
in this active region has been estimated to be above 10$^{33}$ erg,
consistent with a large total photospheric magnetic flux. For the
potential model (see Fig.~\ref{fig:ar0486}), the $\Xi_e \leq 1$
values are located above strong photospheric magnetic field strength
and extend in a volume below 70 Mm. For the \nlff\ field, the model
generates low $\Xi_e$ values high in the corona up to 170 Mm. These
high-altitude, low values of $\Xi_e$ correspond to several pressure
scale-heights. This indicates that  not only is the pressure
scale-height important for this model (as noted for the previous
active regions), but also that the magnetic scale-height is relevant
when the total unsigned magnetic flux on the photosphere is large. In
Fig.~\ref{fig:ar_dist}d, all the distributions of $\Xi_e$ peak at 1;
however, the \nlff\ distribution has a larger number of occurrences
and a narrower width, with 80\% of the values being between 0 and
2.5, compared to the potential distribution which has an extended
tail towards large values of $\Xi_e$. This is  reflected in
Table~\ref{tab:xi}, which shows that the \nlff\ magnetic
configuration produces a larger number of favourable locations for
maser emission in the corona.

\section{Conclusions}
\label{sec:concl}

\begin{table*}
\caption{Percentage of pixels within the computational domain having a $\Xi_e$
included in the given range for the five different examples and for both potential and \nlff\ 
magnetic field models with varying gravity.}    
\label{tab:xi}    
\centering          
\begin{tabular}{cccccc}     
\hline\\[-0.3cm]       
 & Model & $\Xi_e \leq 0.35$ & $0.35 < \Xi_e \leq 1$ & $1 < \Xi_e \leq 2.5$ 
        & $\Xi_e > 2.5$ \\[0.1cm] 
\hline                    
\hline \\[-0.3cm]               
Bipole 
 & Potential & 0.41\% & 1.09\% & 2.76\% & 95.74\%\\
 & \nlff\ & 0.43\% & 1.18\% & 3.06\% & 95.33\% \\[0.1cm]
\hline \\[-0.3cm]               
AR8151 
 & Potential & 0.05\% & 0.73\% & 7.83\% & 91.39\% \\
 & \nlff\ & 0.05\% & 1.33\% & 39.93\% & 58.69\% \\[0.1cm]
\hline \\[-0.3cm]               
AR8210 
 & Potential & 0.10\% & 0.37\% & 0.88\% & 98.65\% \\
 & \nlff\ & 0.10\% & 0.37\% & 0.93\% & 98.60\% \\[0.1cm]
\hline  \\[-0.3cm]              
AR9077 
 & Potential & 2.50\% & 11.48\% & 34.64\% & 51.38\% \\
 & \nlff\ & 2.99\% & 15.53\% & 39.28\% & 42.20\% \\[0.1cm]
\hline \\[-0.3cm]               
AR10486 
 & Potential & 1.74\% & 4.89\% & 14.06\% & 79.31\% \\
 & \nlff\ & 2.48\% & 12.13\% & 48.83\% & 36.56\% \\
\end{tabular}
\end{table*}
                
Using a combination of force-free extrapolation and hydrostatic models, we 
estimate the \ratio\ ratio of coronal magnetic configurations above active regions.  The \ratio\ ratio is
important in order to determine in which regime the plasma can evolve and what kind of electronic plasma waves can propagate and grow. 
The main electronic wave modes are the $O$- and $X$-type for the maser emission, the whistler mode, and the upper hybrid mode (electrostatic mode). The different regimes in which these modes  grow or are suppressed is essentially a function of the \ratio\ ratio. \citet{mel82} have mentioned that the growth of the first and second harmonics of the $X$-mode may be responsible for observed radio and hard X-ray emission in the solar corona. The study of \citet{sha84} has defined intervals of \ratio\ in which the different modes are important. For instance, the first harmonic of the $X$-mode has a maximum growth rate for \ratio $\leq$ 0.35, whilst the first harmonic of the $O$-mode dominates when 0.35$\leq$ \ratio\ $\leq$ 1. The second harmonic of the $X$-mode is dominant for 1$\leq$ \ratio\ $\leq$ 1.45. These values are obtained in the case where the coupling with other wave modes is neglected.
Therefore, it is important to know what values of \ratio\ can be expected in a realistic magnetic field above active regions and thus to check if the electron-cyclotron maser emission can be a viable mechanism in the solar corona. To achieve these goals, we develop a zero-order magnetohydrostatic model that
neglects the coupling between the magnetic field and the plasma \citep{reg08b}. Applying this method to several active regions, we found that
\ratio\ can be less than 0.35 in both potential and \nlff\ field models; however, this is
statistically marginal and thus localised in small coronal volumes.

The two main results obtained from this study are:
\begin{itemize}
\item[1. ]{The smallest values of \ratio\ are located where the magnetic field strength
is the largest at the bottom of the corona over sunspots; this is true whatever the magnetic field model used.}
\item[2. ]{Values of \ratio\ less than 1 can be found high in the corona in the case where
highly twisted flux tubes can be found within the magnetic configuration; this is only obtained in magnetic field models containing electric currents.} 
\end{itemize}

From this new technique for estimating \ratio\ in coronal plasmas, we conclude that the maser instability/emission is a viable mechanism. As mentioned by \citet{mel82}, the second harmonics of the $O$- and $X$-modes are the most possible modes to be observed, whilst the possibility of observing the first harmonic of the $X$-mode is statistically insignificant. In addition, the possible emission is most likely to be localised at the bottom of the corona or at coronal heights where the free magnetic energy is locally increased (for instance, near twisted flux bundles). We also note that both the pressure and magnetic field scale-heights are important in order to describe the variation of the Alfv\'en speed and thus the distribution of \ratio.

The low values of \ratio\ localised high in the corona (see Fig.~\ref{fig:ar8151} and Fig.~\ref{fig:ar0486}) are situated above twisted flux bundles for AR8151 and above sheared arcades for AR10846, and are indeed related to the complex magnetic topology of the fields induced by those structures. To understand better the role that these low \ratio\ values play in the activity of an active region, the next step  is to follow the time evolution of an active region producing observed radio emission consistent with the maser instability.

\begin{acknowledgements}
We would like to thank Alec McKinnon and Don Melrose for useful
discussions and suggestions on this topic. The \nlff\ computations
were performed with the XTRAPOL code developed by T. Amari (supported
by the Ecole Polytechnique, Palaiseau, France and the CNES). The
author acknowledges IDL support provided by STFC as well as the
provision of STFC HPC facilities.
\end{acknowledgements}

\bibliographystyle{aa}
\bibliography{aa_2014_25346}

\begin{thebibliography}{24}
\expandafter\ifx\csname natexlab\endcsname\relax\def\natexlab#1{#1}\fi

\bibitem[{{Amari} {et~al.}(1997){Amari}, {Aly}, {Luciani}, {Boulmezaoud}, \&
  {Mikic}}]{ama97}
{Amari}, T., {Aly}, J.~J., {Luciani}, J.~F., {Boulmezaoud}, T.~Z., \& {Mikic},
  Z. 1997, Solar Phys., 174, 129

\bibitem[{{Amari} {et~al.}(1999){Amari}, {Boulmezaoud}, \& {Mikic}}]{ama99b}
{Amari}, T., {Boulmezaoud}, T.~Z., \& {Mikic}, Z. 1999, A\&A, 350, 1051

\bibitem[{{Grad} \& {Rubin}(1958)}]{gra58}
{Grad}, H. \& {Rubin}, H. 1958, in {Proc. 2nd Int. Conf. on Peaceful Uses of
  Atomic Energy, Geneva, UN}, Vol.~31, 190

\bibitem[{{Holman} {et~al.}(1980){Holman}, {Eichler}, \& {Kundu}}]{hol80}
{Holman}, G.~D., {Eichler}, D., \& {Kundu}, M.~R. 1980, in IAU Symposium,
  Vol.~86, Radio Physics of the Sun, ed. M.~R. {Kundu} \& T.~E. {Gergely},
  457--459

\bibitem[{{Kosovichev} \& {Zharkova}(2001)}]{kos01}
{Kosovichev}, A.~G. \& {Zharkova}, V.~V. 2001, \apjl, 550, L105

\bibitem[{{Lee} {et~al.}(2013){Lee}, {Yi}, {Lim}, {Kim}, {Seough}, \&
  {Yoon}}]{lee13}
{Lee}, S.-Y., {Yi}, S., {Lim}, D., {et~al.} 2013, Journal of Geophysical
  Research (Space Physics), 118, 7036

\bibitem[{{Mandrini} {et~al.}(2006){Mandrini}, {Demoulin}, {Schmieder},
  {Deluca}, {Pariat}, \& {Uddin}}]{man06}
{Mandrini}, C.~H., {Demoulin}, P., {Schmieder}, B., {et~al.} 2006, \solphys,
  238, 293

\bibitem[{{Melrose} \& {Dulk}(1982)}]{mel82}
{Melrose}, D.~B. \& {Dulk}, G.~A. 1982, \apj, 259, 844

\bibitem[{{Melrose} {et~al.}(1984){Melrose}, {Dulk}, \& {Hewitt}}]{mel84}
{Melrose}, D.~B., {Dulk}, G.~A., \& {Hewitt}, R.~G. 1984, \jgr, 89, 897

\bibitem[{{R{\'e}gnier}(2012)}]{reg12}
{R{\'e}gnier}, S. 2012, Solar Phys., 277, 131

\bibitem[{{R{\'e}gnier}(2013)}]{reg13a}
{R{\'e}gnier}, S. 2013, Solar Phys., 288, 481

\bibitem[{{R{\'e}gnier} \& {Amari}(2004)}]{reg04}
{R{\'e}gnier}, S. \& {Amari}, T. 2004, A\&A, 425, 345

\bibitem[{{R{\'e}gnier} {et~al.}(2002){R{\'e}gnier}, {Amari}, \& {Kersal{\'
  e}}}]{reg02}
{R{\'e}gnier}, S., {Amari}, T., \& {Kersal{\' e}}, E. 2002, A\&A, 392, 1119

\bibitem[{{R{\'e}gnier} \& {Canfield}(2006)}]{reg06}
{R{\'e}gnier}, S. \& {Canfield}, R.~C. 2006, A\&A, 451, 319

\bibitem[{{R{\'e}gnier} {et~al.}(2005){R{\'e}gnier}, {Fleck}, {Abramenko}, \&
  {Zhang}}]{reg05c}
{R{\'e}gnier}, S., {Fleck}, B., {Abramenko}, V., \& {Zhang}, H.~Q. 2005, in
  ESA-SP 596: Chromospheric and coronal magnetic field meeting, Lindau, Germany

\bibitem[{{R{\'e}gnier} {et~al.}(2008){R{\'e}gnier}, {Priest}, \&
  {Hood}}]{reg08b}
{R{\'e}gnier}, S., {Priest}, E.~R., \& {Hood}, A.~W. 2008, A\&A, 491, 297

\bibitem[{{Sharma} \& {Vlahos}(1984)}]{sha84}
{Sharma}, R.~R. \& {Vlahos}, L. 1984, \apj, 280, 405

\bibitem[{{Tang} \& {Wu}(2009)}]{tan09}
{Tang}, J.~F. \& {Wu}, D.~J. 2009, \aap, 493, 623

\bibitem[{{Treumann}(2006)}]{tre06}
{Treumann}, R.~A. 2006, \aapr, 13, 229

\bibitem[{{Vocks} \& {Mann}(2004)}]{voc04}
{Vocks}, C. \& {Mann}, G. 2004, \aap, 419, 763

\bibitem[{{Wang} {et~al.}(2005){Wang}, {Liu}, {Deng}, \& {Zhang}}]{wan04}
{Wang}, H., {Liu}, C., {Deng}, Y., \& {Zhang}, H. 2005, \apj, 627, 1031

\bibitem[{{Wiegelmann} \& {Sakurai}(2012)}]{wie12}
{Wiegelmann}, T. \& {Sakurai}, T. 2012, Living Reviews in Solar Physics, 9, 5

\bibitem[{{Winglee}(1985)}]{win85}
{Winglee}, R.~M. 1985, \jgr, 90, 9663

\bibitem[{{Zuccarello} {et~al.}(2009){Zuccarello}, {Romano}, {Farnik},
  {Karlicky}, {Contarino}, {Battiato}, {Guglielmino}, {Comparato}, \&
  {Ugarte-Urra}}]{zuc09}
{Zuccarello}, F., {Romano}, P., {Farnik}, F., {et~al.} 2009, \aap, 493, 629

\end{thebibliography}

\appendix
\section{Maser emission and twisted flux tubes in a constant gravity field} 

\begin{figure}[!ht]
\centering
\includegraphics[width=.7\linewidth, bb= 60 0 510 340]
       {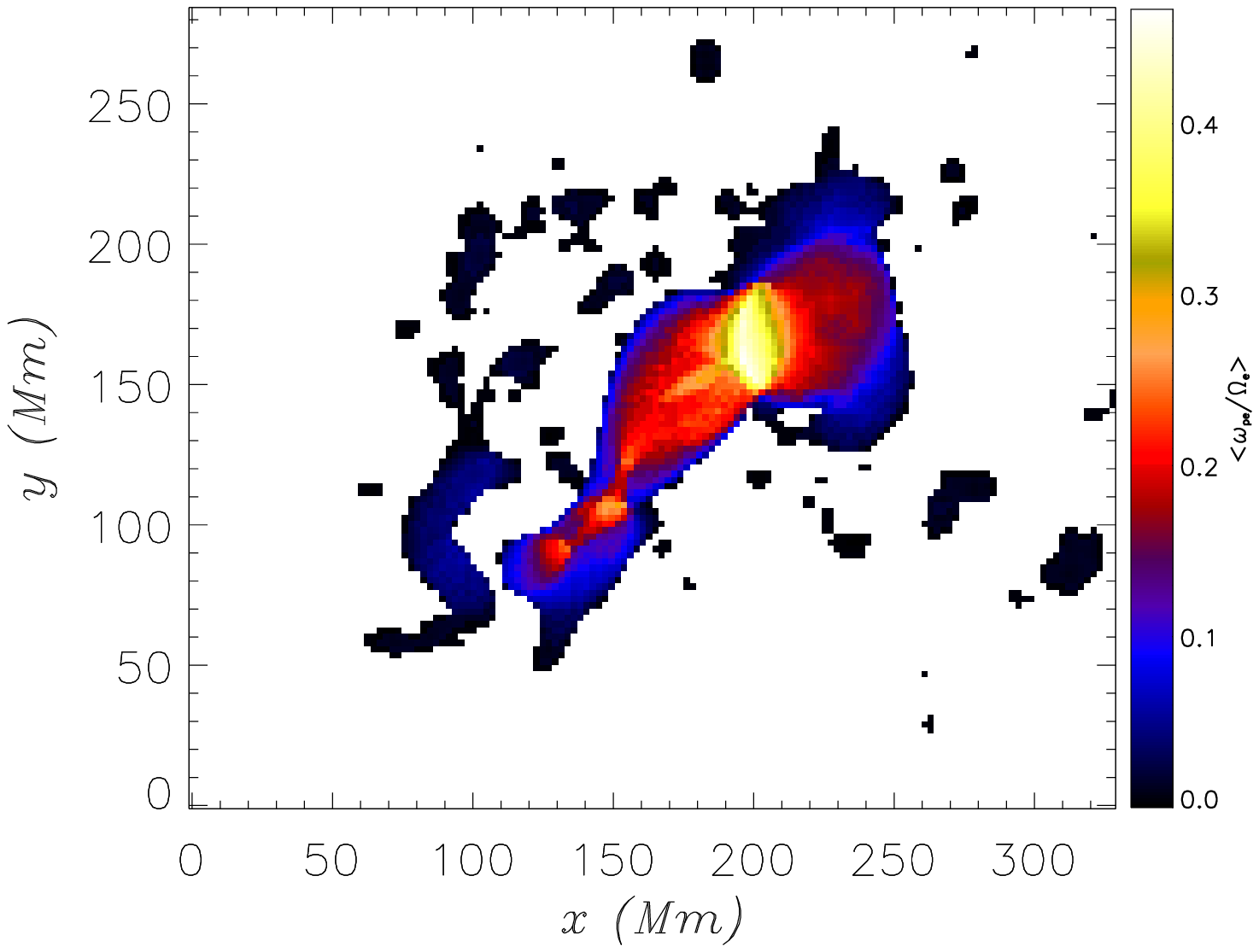}
\includegraphics[width=.7\linewidth, bb= 20 0 480 300]
       {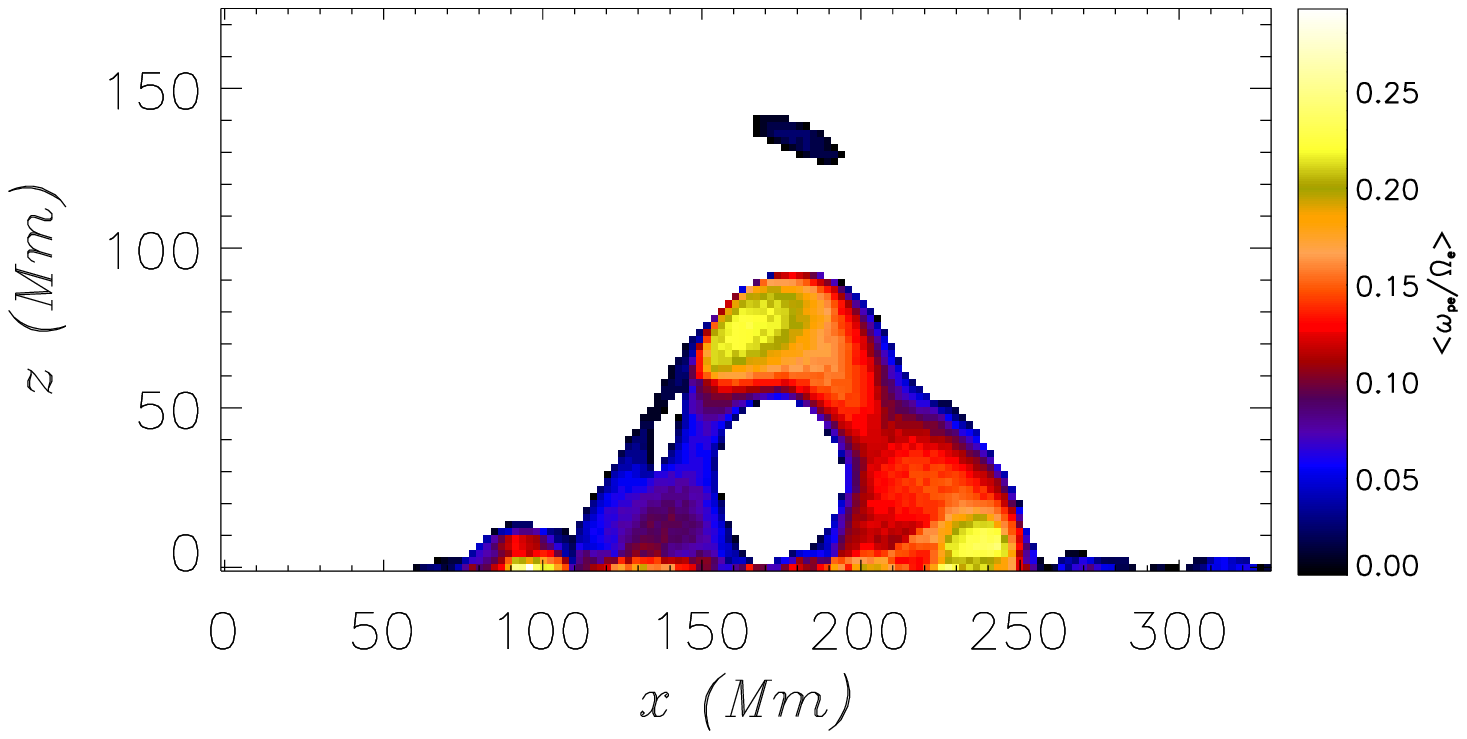}
\caption{Same as Fig.~\ref{fig:rng10} for AR8210}
\label{fig:ar8210_nlff}
\end{figure}

For the sake of completeness, we present the $\Xi_e$ values derived for AR8151 in the case of a \nlff\ field with constant gravity, which exhibits a peculiar feature. For the constant gravity model, the plasma pressure and density, and the Alfv\'en speed are thus given by
\begin{equation}
\centering
\left\{ \begin{array}{c l}
p(z) = p_0 \exp{\left(- \frac{z - z_0}{H} \right)}, & \quad \textrm{and} \\[0.2cm]
\rho(z) = \rho_0 \exp{\left(- \frac{z - z_0}{H} \right)} & \quad \textrm{for $g = g_0$}\\
\end{array}
\right.  
\end{equation}
and
\begin{equation}
v_A(x, y, z) = \frac{B(x, y, z)}{\sqrt{\mu_0 \rho_0}} \exp{\left( \frac{z - z_0}{2H}
\right)},
\label{eq:va}
\end{equation}
where $H = k_BT/(\tilde{\mu}m_pg_0)$ is the pressure scale-height ($k_B = 1.38~10^{-23}$ J~K$^{-1}$, $\tilde{\mu} = 0.6$ for a fully ionised coronal plasma, $m_p = 1.67~10^{-27}$ kg, and $g_0 = 274$
m~s$^{-2}$), and $p_0$ and $\rho_0$ are characteristic values of the plasma pressure and density at $z_0$.  In Fig.~\ref{fig:ar8210_nlff}, an isolated blob of low $\Xi_e$ values appears at a height of about 130 Mm. Comparing with Fig.~\ref{fig:ar8210}, the blob is not present in the \nlff\ model with varying gravity. The blob is located above the highly twisted flux tube that has been defined in \cite{reg04}. A possible explanation for the appearance of the blob in this particular model is that the existence of the twisted flux tube changes the decay rate of the magnetic field strength (i.e. the magnetic field scale-height), whilst the density still follows the hydrostatic decay with height. The blob does not appear when the gravity  varies with height because the density  drops faster than the magnetic field strength as emphasised in Fig. 7 of \cite{reg08b}. The blob is thus an artefact of this particular model; however, its existence emphasises the importance of understanding the variation of both the magnetic and pressure scale-heights within the corona. 

\end{document}